\documentclass[usenatbib,12pt,preprint]{aastex6}
\usepackage{natbib}

\slugcomment{Version: \today}

\usepackage{color}

\begin{document}
\title{The JCMT Gould Belt Survey: A First Look at IC\,5146}
  \author{
  D. Johnstone\altaffilmark{1,2},
  S. Ciccone\altaffilmark{1,3}, 
  H. Kirk\altaffilmark{1}, 
  S. Mairs\altaffilmark{1,2}, 
  J. Buckle\altaffilmark{4,5}, 
  D.S. Berry\altaffilmark{6},  
  H. Broekhoven-Fiene\altaffilmark{1,2},  
  M.J. Currie\altaffilmark{6}, 
  J. Hatchell\altaffilmark{7}, 
  T. Jenness\altaffilmark{8},  
  J.C. Mottram\altaffilmark{9,10}, 
  K. Pattle\altaffilmark{11}, 
  S. Tisi\altaffilmark{12},
  J. Di Francesco\altaffilmark{1,2}, 
  M.R. Hogerheijde\altaffilmark{9},  
  D. Ward-Thompson\altaffilmark{11}, 
  P. Bastien\altaffilmark{13},  
  D. Bresnahan\altaffilmark{11},  
  H. Butner\altaffilmark{14}, 
  M. Chen\altaffilmark{1,2},  
  A. Chrysostomou\altaffilmark{15}, 
  S. Coud\'e\altaffilmark{13},  
  C.J. Davis\altaffilmark{16},  
  E. Drabek-Maunder\altaffilmark{17}, 
  A. Duarte-Cabral\altaffilmark{7}, 
  M. Fich\altaffilmark{12}, 
  J. Fiege\altaffilmark{18}, 
  P. Friberg\altaffilmark{6},  
  R. Friesen\altaffilmark{19}, 
  G.A. Fuller\altaffilmark{20},  
  S. Graves\altaffilmark{6}, 
  J. Greaves\altaffilmark{21}, 
  J. Gregson\altaffilmark{22,23},
  W. Holland\altaffilmark{24,25},  
  G. Joncas\altaffilmark{26}, 
  J.M. Kirk\altaffilmark{11}, 
  L.B.G. Knee\altaffilmark{1}, 
  K. Marsh\altaffilmark{27},  
  B.C. Matthews\altaffilmark{1,2},
  G. Moriarty-Schieven\altaffilmark{1}, 
  C. Mowat\altaffilmark{7}, 
  D. Nutter\altaffilmark{27}, 
  J.E. Pineda\altaffilmark{20,28,29}, 
  C. Salji\altaffilmark{4,5},   
  J. Rawlings\altaffilmark{30},  
  J. Richer\altaffilmark{4,5}, 
  D. Robertson\altaffilmark{3}, 
  E. Rosolowsky\altaffilmark{31}, 
  D. Rumble\altaffilmark{7}, 
  S. Sadavoy\altaffilmark{10},  
  H. Thomas\altaffilmark{6}, 
  N. Tothill\altaffilmark{32}, 
  S. Viti\altaffilmark{30}, 
  G.J. White\altaffilmark{22,23}, 
  J. Wouterloot\altaffilmark{6}, 
  J. Yates\altaffilmark{30}, 
  M. Zhu\altaffilmark{33}
}

\altaffiltext{1}{NRC Herzberg Astronomy and Astrophysics, 5071 West Saanich Rd, Victoria, BC, V9E 2E7, Canada}
\altaffiltext{2}{Department of Physics and Astronomy, University of Victoria, Victoria, BC, V8P 1A1, Canada}
\altaffiltext{3}{Department of Physics and Astronomy, McMaster University, Hamilton, ON, L8S 4M1, Canada}
\altaffiltext{4}{Astrophysics Group, Cavendish Laboratory, J J Thomson Avenue, Cambridge, CB3 0HE, UK}
\altaffiltext{5}{Kavli Institute for Cosmology, Institute of Astronomy, University of Cambridge, Madingley Road, Cambridge, CB3 0HA, UK}
\altaffiltext{6}{East Asian Observatory, 660 North A`oh\={o}k\={u} Place, University Park, Hilo, Hawaii 96720, USA}
\altaffiltext{7}{Physics and Astronomy, University of Exeter, Stocker Road, Exeter EX4 4QL, UK}
\altaffiltext{8}{Large Synoptic Survey Telescope Project Office, 933 N. Cherry Ave, Tucson, Arizona 85721, USA}
\altaffiltext{9}{Leiden Observatory, Leiden University, PO Box 9513, 2300 RA Leiden, The Netherlands}
\altaffiltext{10}{Max Planck Institute for Astronomy, K{\"o}nigstuhl 17, D-69117 Heidelberg, Germany}
\altaffiltext{11}{Jeremiah Horrocks Institute, University of Central Lancashire, Preston, Lancashire, PR1 2HE, UK}
\altaffiltext{12}{Department of Physics and Astronomy, University of Waterloo, Waterloo, Ontario, N2L 3G1, Canada}
\altaffiltext{13}{Universit\'e de Montr\'eal, Centre de Recherche en Astrophysique du Qu\'ebec et d\'epartement de physique, C.P. 6128, succ.~centre-ville, Montr\'eal, QC, H3C 3J7, Canada}
\altaffiltext{14}{James Madison University, Harrisonburg, Virginia 22807, USA}
\altaffiltext{15}{School of Physics, Astronomy \& Mathematics, University of Hertfordshire, College Lane, Hatfield, Herts, AL10 9AB, UK}
\altaffiltext{16}{Astrophysics Research Institute, Liverpool John Moores University, Egerton Warf, Birkenhead, CH41 1LD, UK}
\altaffiltext{17}{Imperial College London, Blackett Laboratory, Prince Consort Rd, London SW7 2BB, UK}
\altaffiltext{18}{Dept of Physics \& Astronomy, University of Manitoba, Winnipeg, Manitoba, R3T 2N2 Canada}
\altaffiltext{19}{Dunlap Institute for Astronomy \& Astrophysics, University of Toronto, 50 St. George St., Toronto ON M5S 3H4 Canada}
\altaffiltext{20}{Jodrell Bank Centre for Astrophysics, Alan Turing Building, School of Physics and Astronomy, University of Manchester, Oxford Road, Manchester, M13 9PL, UK}
\altaffiltext{21}{Physics \& Astronomy, University of St Andrews, North Haugh, St Andrews, Fife KY16 9SS, UK}
\altaffiltext{22}{Dept. of Physical Sciences, The Open University, Milton Keynes MK7 6AA, UK}
\altaffiltext{23}{The Rutherford Appleton Laboratory, Chilton, Didcot, OX11 0NL, UK}
\altaffiltext{24}{UK Astronomy Technology Centre, Royal Observatory, Blackford Hill, Edinburgh EH9 3HJ, UK}
\altaffiltext{25}{Institute for Astronomy, Royal Observatory, University of Edinburgh, Blackford Hill, Edinburgh EH9 3HJ, UK}
\altaffiltext{26}{Centre de recherche en astrophysique du Qu\'ebec et D\'epartement de physique, de g\'enie physique et d'optique, Universit\'e Laval, 1045 avenue de la m\'edecine, Qu\'ebec, G1V 0A6, Canada}
\altaffiltext{27}{School of Physics and Astronomy, Cardiff University, The Parade, Cardiff, CF24 3AA, UK}
\altaffiltext{28}{European Southern Observatory (ESO), Garching, Germany}
\altaffiltext{29}{Max Planck Institute for Extraterrestrial Physics, Giessenbachstrasse 1, 85748 Garching, Germany}
\altaffiltext{30}{Department of Physics and Astronomy, UCL, Gower St, London, WC1E 6BT, UK}
\altaffiltext{31}{Department of Physics, University of Alberta, Edmonton, AB T6G 2E1, Canada}
\altaffiltext{32}{University of Western Sydney, Locked Bag 1797, Penrith NSW 2751, Australia}
\altaffiltext{33}{National Astronomical Observatory of China, 20A Datun Road, Chaoyang District, Beijing 100012, China}

\begin{abstract}
We present 450\,$\mu$m and 850\,$\mu$m submillimetre continuum observations of the IC\,5146 star-forming region taken as part of the JCMT Gould Belt Survey.  We investigate the location of bright submillimetre (clumped) emission with the larger-scale  molecular cloud through comparison with extinction maps, 
and find that these denser structures correlate with higher cloud column density. Ninety-six individual submillimetre clumps are identified using FellWalker and their physical properties are examined. These clumps are found to be relatively massive, ranging from 0.5\,$M_\odot$ to 116\,$M_\odot$ with a mean mass of 8\,$M_\odot$ and a median mass of 3.7\,$M_\odot$. A stability analysis for the clumps suggest that the majority are (thermally) Jeans stable, with $M/M_J < 1$. We further compare the locations of known protostars with the observed submillimetre emission, finding that younger protostars, i.e., Class 0 and I sources, are strongly correlated with submillimetre peaks and that the clumps with protostars are among the most Jeans unstable. Finally, we contrast the evolutionary conditions in the two major star-forming regions within IC\,5146: the young cluster associated with the Cocoon Nebula and the more distributed star formation associated with the Northern Streamer filaments.  
The Cocoon Nebula appears to have converted a higher fraction of its mass into dense clumps and protostars, the clumps are more likely to be Jeans unstable, and a larger fraction of these remaining clumps contain embedded protostars. The Northern Streamer, however, has a larger number of clumps in total and a larger fraction of the known protostars are still embedded within these clumps. 
\end{abstract}

\keywords{stars:formation - stars:protostars - ISM:structure - ISM:clouds - submillimetre:general - submillimetre:ISM}

\section{Introduction}
\label{sec:intro}
The Gould Belt Legacy Survey \citep[GBS;][]{wtea07} conducted
with the James Clerk Maxwell Telescope (JCMT) extensively observed
many nearby star-forming regions, tracing the very earliest stages
of star formation at 450\,$\mu$m and 850\,$\mu$m with the Submillimetre 
Common-User Bolometer Array 2 \citep[SCUBA-2;][]{hea06}. 
This imaging covered $50$ square degrees of nearby clouds within the Gould Belt,
including well-known regions such as 
Auriga \citep{bea16},
Ophiuchus \citep{pea15}, 
Orion \citep{sea15a, sea15b, kea16a, Kirk16b, mea16, Lane16},
Perseus \citep{sea13, Hatchell13, cea16}, 
Serpens~MWC~297 \citep{rea15}, 
Taurus \citep{bea15,wtea16}, 
and
W40 \citep{rea16}.
Among these targets, the GBS survey covered approximately $2.5$ square degrees 
of the  IC\,5146 star-forming region. 

The molecular cloud, IC\,5146, is both a reflection nebula and an HII
region surrounding the B0 V Star BD+46$^{\circ}$ 3474 \citep[][see their Figures 1 and 2]{hr08}.
IC\,5146 is comprised of two notable features: the first being the Cocoon Nebula which is
a bright core nebula located within a bulbous dark cloud
at the end of a long filamentary second feature, the Northern Streamer,
extending northwest from the Cocoon Nebula. These two features display
distinctly different properties, clustered versus distributed young stars, and therefore present an ideal laboratory for
investigating the range of star formation processes within a single cloud.

In this paper, we take a first look at the submillimetre continuum emission within IC\,5146, 
concentrating on the distribution of dense gas and dust within the cloud and its relation with 
on-going star formation. In \S\ref{sec:IC}, we provide background information on IC\,5146 and 
its two main features. The new SCUBA-2 observations along with estimates of the total cloud 
column density and protostellar content are discussed in \S\ref{sec:obs}.  These observations 
are analyzed in \S\ref{sec:anal}, starting from the largest physical scales and zooming in 
through the submillimetre clumps to the individual young stellar objects. Section~\ref{sec:disc} 
adds context to the analysis, contrasting star formation in the Cocoon Nebula and the Northern 
Streamer filaments, and the results are summarized in \S\ref{sec:conc}.

\section{IC\,5146}
\label{sec:IC}

There are varying estimates of the distance to IC\,5146. 
Initially, a distance of 1000\,pc was determined by \cite{w59} 
using photoelectric star observations.
A distance of 460~pc was later derived by \cite{lal99} using deep
near-infrared (HK) imaging observations to compare the number of foreground
stars to those expected from galactic models. \cite{hr08}
adopt a distance of $1200 \pm 180$\,pc in their review based on the
work of \cite{hd02} that used spectroscopic distances to
late-B stars and two different main-sequence calibrations. 
An estimation of 1200~pc is quite high in comparison
to the Lada estimation but nearer to initial stellar distances measured
by \cite{w59} and \cite{e78}. For their {\it Spitzer} Space Telescope analysis of
IC\,5146, \cite{hea08} re-evaluated the photometric 
distance using a modern ZAMS calibrator, the Orion Nebula Cluster (ONC), 
and several photometric methods for different members of IC\,5146. A distance
of $950 \pm 80$\,pc was determined by their analysis and, for consistency with that paper, 
we use $950\,$pc throughout the rest of this work.

IC\,5146 consists of many distinct populations across two main features,
the Cocoon Nebula and the Northern Streamer.
Optical and near-infrared identified objects include approximately 20 variable stars,
40 faint stars above the main sequence that are the members of a
population of young pre-main sequence stars, 100 H$\alpha$-emission
stars \citep{hd02}, 110 $M>1.2M_{\odot}$ stars \citep{fo84}, 
and 200 candidate young stellar objects \citep{hea08}. 
Using WISE data in combination with existing {\it Spitzer} and 2MASS observations,
\citet{Nunes16} identify new candidate YSOs in the extended IC~5146 region, including
five protostellar clusters (with $\sim 20 - 50$ members each) in the area around the 
Northern Streamer, as well as a total of $\sim$160 protostars within the cluster
at the centre of the Cocoon Nebula.
The more-evolved stars of the Cocoon Nebula complex are thought to be spatially co-distant
with the younger stars still entangled
within the current dense gas. Two discrepancies to this claim
are the distances determined for BD+46$^{\circ}$ 3471 and BD+46$^{\circ}$ 3474 as 355\,pc and
400 pc respectively \citep{hea08}. These distances, as well
as the \cite{lal99} distance of 460\,pc, remain problematic because
at approximately 400\,pc the ages of the K/M-type T Tauri stars in the cluster 
jump from 0.2 Myr to 15 Myr \citep{hea08}. The later isochronal age is inconsistent
with the fact that IC\,5146 has a high number of accreting pre-main
sequence stars, as well as the appearance of nebulosity observed in
the region. Indeed, this evidence of on-going star formation argues for an
age of less than a few Myrs for the region. There is also strong circumstantial 
evidence that the two main components of IC\,5146, the Cocoon Nebula and the
Northern Streamer, are co-distant. First, 
the Cocoon Nebula appears to be connected to a long filament with the Northern
Streamer forming at the opposite end. Second, the molecular gas associated
with the Cocoon Nebula has velocities consistent
with those seen in the rest of the IC\,5146 region \citep{dea93}.

The total H$_{2}$ molecular mass of the central IC\,5146 cloud complex has been estimated
to be approximately $4000\,M_{\odot}$, similar to that of the Taurus
complex, as measured by \cite{dea92, dea93} using $^{12}$CO$(J=1-0)$,
$^{13}$CO$(J=1-0)$, and C$^{18}$O$(J=1-0)$ emission lines. The total mass 
estimates of HI and HII, respectively, are approximately $670\, M_{\odot}$ and $4.5\: M_{\odot}$
as determined by \cite{s95} or $445\pm105\: M_{\odot}$ and $9.8\pm1\: M_{\odot}$
as determined by \cite{ri82}. All of these mass estimates were made assuming a distance to the
cloud of approximately 1\,kpc. The median age for the cluster, based on isochrone fits to the optical 
and near-infrared pre-main sequence stars,
is approximately 1.0 Myr \citep{hd02, hr08}, in rough agreement with the young ages
determined for the K/M-type T Tauri stars.

\subsection{The Cocoon Nebula}
\label{sec:IC:cocoon}

The molecular cloud structure southeast of the Northern Streamer is
noticeably bright and active due in part to the presence of BD+46$^{\circ}$
3474 and the high-luminosity variable star V1735 Cyg from which 
X-ray emission has been detected \citep{sea09}.

The Cocoon is a more-evolved HII star-forming region relative to younger complexes
found in the Orion or Ophiuchus clouds, showing decreased small-scale
structure due to density inhomogeneities expanding into more diffuse
surroundings \citep{i77, wea84}. The dust has dispersed into
a mottled structure, as evidenced from scattered light, 
suggesting that the collective activity of the stars in this 
region has blown out material from the centre.
There is further evidence of dispersal due to the presence of BD+46$^{\circ}$
3474 itself, which is the most massive star in IC\,5146 and is responsible
for the observed gas and dust emission surrounding the HII region
\citep{wea84}. Cluster stars appear to have formed in a dense foreground
section of the molecular cloud. BD+46$^{\circ}$ 3474, however, has evacuated a blister
cavity out of which gas and dust are now flowing through a funnel-shaped
volume and dissipating the cloud region \citep{hd02}.

More than 200 young stellar object (YSO) candidates have been identified 
throughout IC\,5146 using 
observations from the {\it Spitzer} Space Telescope, and
the population near the Cocoon Nebula is more evolved, based on spectral observations,
when compared to the younger YSOs found in the Northern Streamer \citep{hea08}. 
High-velocity CO outflows were identified around many
protostellar candidates, based on the IRAS Point Source Catalog
\citep{dea93, dea01}. Thus, this region of the molecular cloud
remains active, exhibiting an extended period of development.

\subsection{The Northern Streamer}
\label{sec:IC:streamer}

The Northern Streamer is comprised of a network of near-parallel filaments in 
which star formation is occurring. Twenty-seven filaments were
identified using {\it Herschel} continuum data and traced throughout the region \citep{aea11}. 
The observed substructure within these filaments suggests that they are the primary birth sites of prestellar cores 
\citep{d12,poea13}. 
We identify both cores and YSOs along the filamentary sections of
the streamer (see \S \ref{sec:anal:clumps} and \ref{sec:anal:yso}), 
supporting the notion 
that here large-scale filament morphology plays a role in the production of stars.

In this paper, we do not characterize or identify any singular
filamentary structures by a modelling algorithm but we do study and analyze
the general morphology of the filamentary and clump structures seen
in the streamer.  Recently, \cite{pea11, pea12} showed that filamentary
geometry at this scale is the most favourable scenario in which
isothermal perturbations grow before global collapse overwhelms
the region dynamics, with the filamentary ends most likely 
to collapse first. In some numerical simulations,
nearly all cores that are detected are associated with filaments and most of these eventually
form protostars \citep[e.g.,][]{mea14}. 
Observations also suggest a strong connection between filaments and cores:
various {\it Herschel} analyses have found that between two-thirds and three-quarters of
cores are located along filaments \citep{Polychroni13,Schisano14,Konyves15}.
Collapse patterns in the Cocoon Nebula
and Northern Streamer provide an opportunity to strengthen
our understanding about the processes playing a pivotal role 
in fragmenting molecular clouds.

\section{Observations and Data Reduction}
\label{sec:obs}

\subsection{SCUBA-2}
\label{sec:obs:S2}

\begin{figure}[htb]
\includegraphics[scale=0.8]{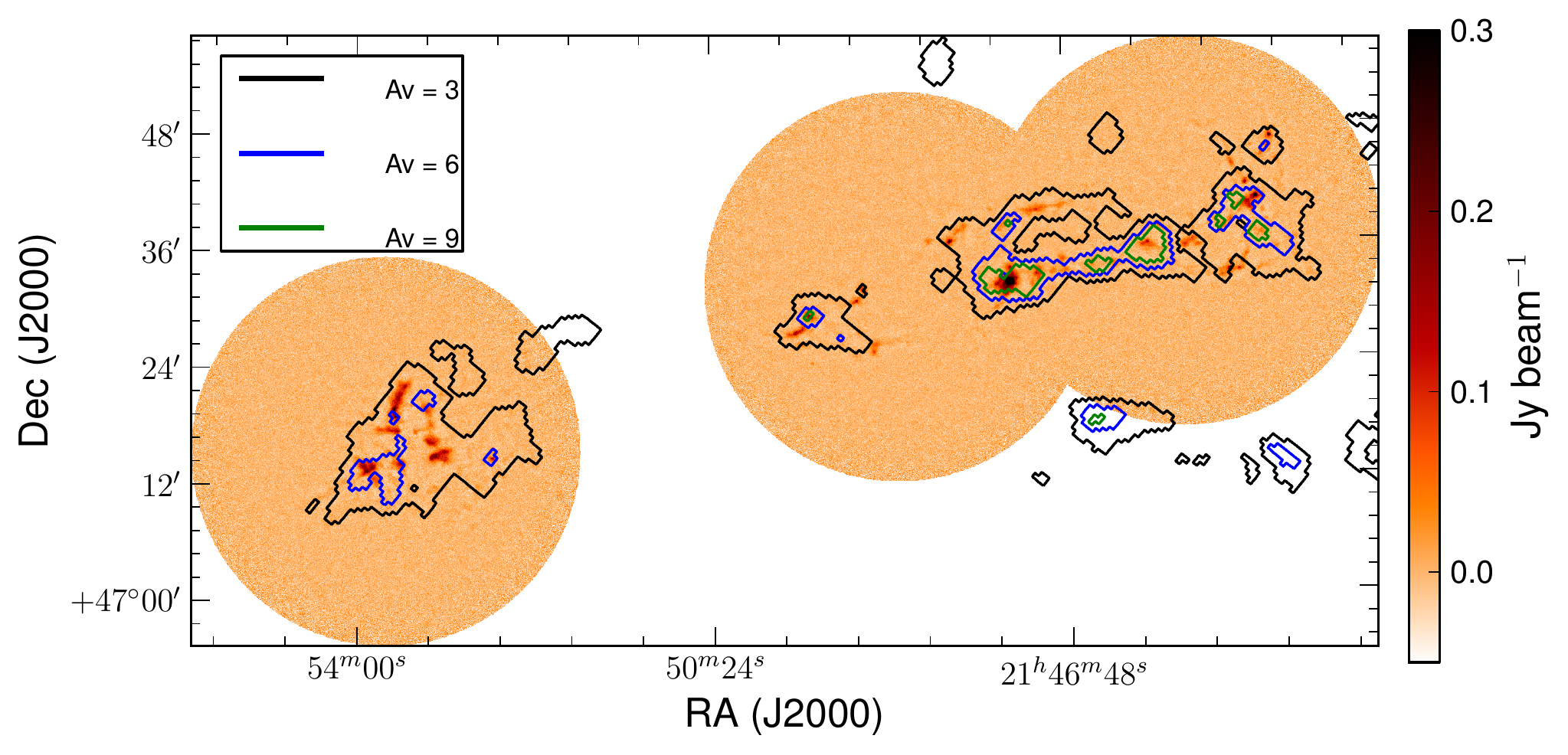}
\caption{
IC\,5146 observed at 850\,$\mu$m with SCUBA-2. Also shown are the $A_V=3$, 6, and 9 contours from the 2MASS-based extinction map (\S \ref{sec:obs:ext}). The {\it Spitzer} coverage of the region (\S \ref{sec:obs:yso}) is almost identical to the SCUBA-2 coverage shown here.
}
\label{fig:full}
\end{figure}

\begin{figure}
\includegraphics[scale=0.4]{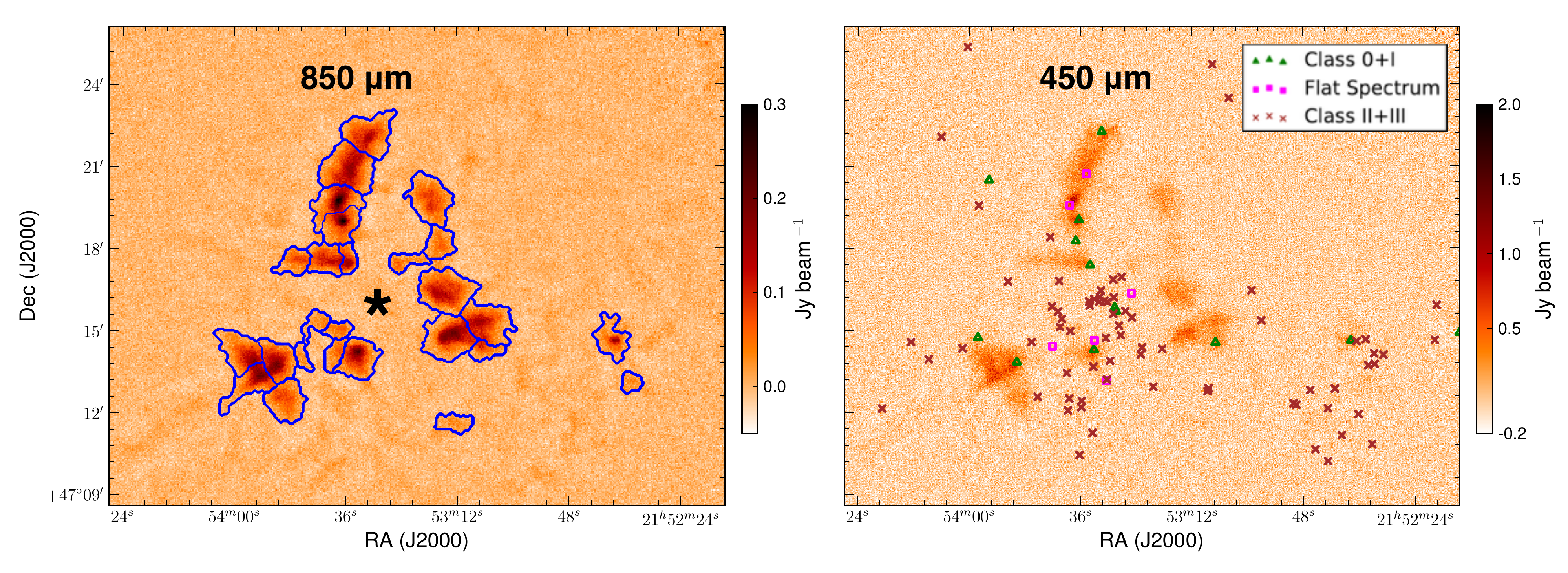}
\caption{
IC\,5146 Cocoon Nebula. Left panel shows 850\,$\mu$m dust continuum emission. Right panel shows 450\,$\mu$m dust emission. Overlaid on the 850\,$\mu$m map are contours denoting the boundaries of the clumps identified in this paper (\S 4.2) while on the 450\,$\mu$m map 
the locations of the {\it Spitzer}-classified YSOs, by type (see legend), are provided (\S 4.3). In this region there are 13 Class 0/I, 6 Flat, 65 Class II, and 9 Class III YSOs.
The black star in the left panel shows the location of the B0 V Star BD+46$^{\circ}$ 3474 \protect\citep{hr08}.}

\label{fig:cocoon}
\end{figure}

\begin{figure}[htb]
\includegraphics[scale=0.4]{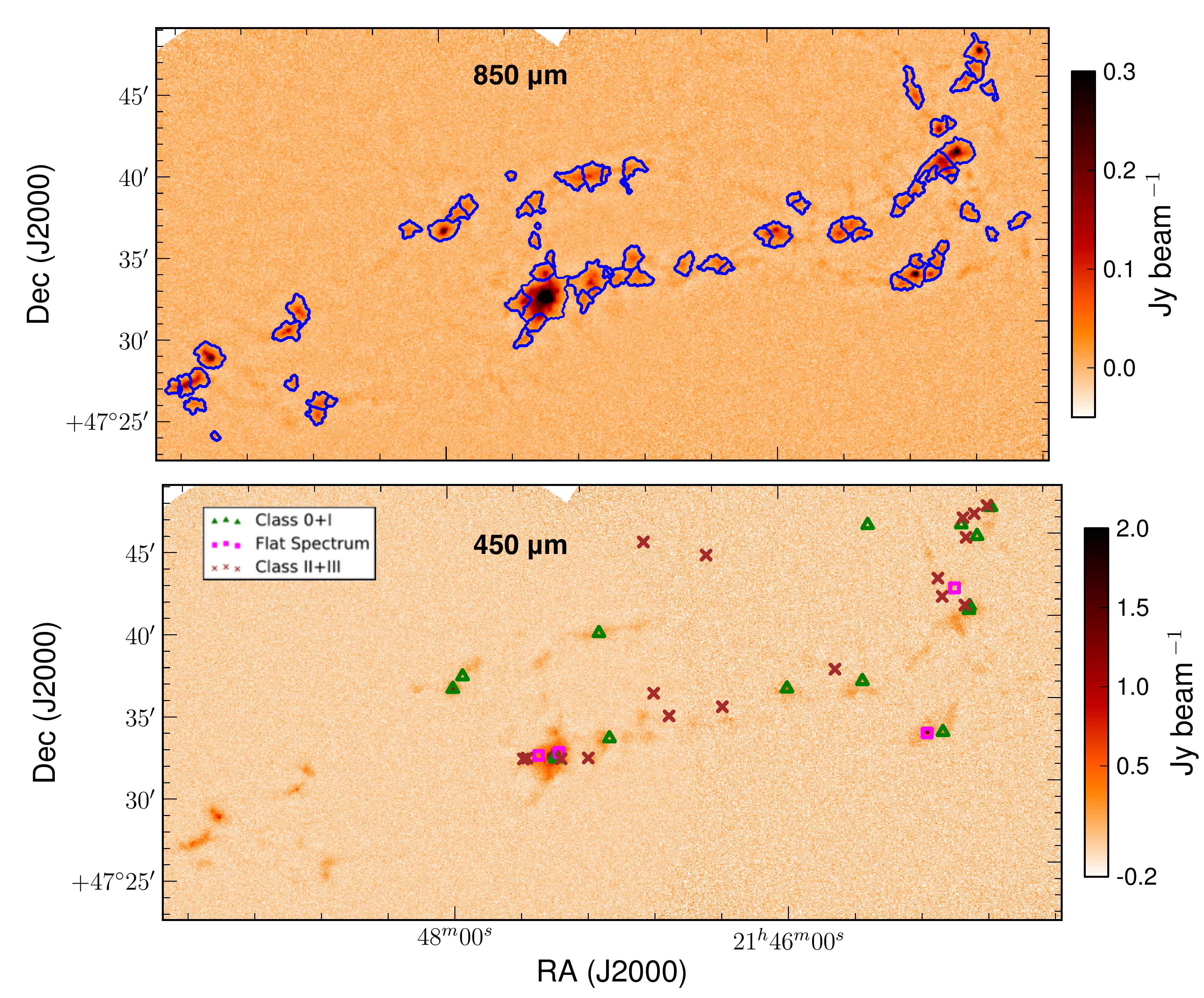}
\caption{
IC\,5146 Northern Streamer. Upper panel shows 850\,$\mu$m dust continuum emission. Lower panel shows 450\,$\mu$m dust emission. Overlaid on the 850\,$\mu$m map are contours denoting the boundaries of the clumps identified in this paper (\S 4.2) while on the 450\,$\mu$m map 
the locations of the YSOs, by type (see legend), are provided (\S 4.3). In this region there are 15 Class 0/I, 4 Flat, 14 Class II, and 6 Class III YSOs.
}
\label{fig:filament}
\end{figure}

IC\,5146 was observed with SCUBA-2 \citep{hea13} at 450\,$\mu$m and 850\,$\mu$m 
simultaneously as part of the 
JCMT Gould Belt Survey \citep[GBS,][]{wtea07}. 
The SCUBA-2 observations were obtained between July 8, 2012 and July 14, 2013.  These data 
were observed as three fully sampled 30\arcmin\ diameter circular regions using the PONG 1800 mode 
\citep{kea10}.  Each area of sky was observed six times.  Neighbouring fields 
were set up to overlap slightly to create a more uniform
noise in the final mosaic.   Details for these observations are provided in Table \ref{tab:noise}, and 
the full IC\,5146 region, observed at 850\,$\mu$m, is shown in Figure \ref{fig:full}.
Figures \ref{fig:cocoon} and \ref{fig:filament} focus on the areas of 450\,$\mu$m and 850\,$\mu$m emission within
Cocoon Nebula and the Northern Streamer, respectively. 

\begin{deluxetable}{llcccccc}
\tablecolumns{8}
\tablewidth{0pc}
\tablecaption{Noise values within the SCUBA-2 maps\label{tab:noise}}
\tablehead{
\colhead{Region} &
\colhead{Name\tablenotemark{a}} & 
\colhead{R.A.\tablenotemark{b}}  & 
\colhead{Dec.\tablenotemark{b}}  & 
\colhead{$\sigma_{\rm 850}$\tablenotemark{c}}  & 
\colhead{$\sigma_{\rm 450}$\tablenotemark{c}}  &
\colhead{$\sigma_{\rm 850}$\tablenotemark{d}}  & 
\colhead{$\sigma_{\rm 450}$\tablenotemark{d}}\\
\colhead{}&
\colhead{}&
\colhead{(J2000.0)} &
\colhead{(J2000.0)} &
\multicolumn{2}{c}{(mJy\,arcsec$^{-2}$)} &
\multicolumn{2}{c}{(mJy\,bm$^{-1}$)} 
}
\startdata
Coccon Nebula& IC5146-H2 & 21:53:45 & 47:15:22 & 0.050& 1.4& 3.7 & 67\\
Northern Streamer& IC5146-E & 21:48:31 & 47:31:48 & 0.045& 1.1& 3.3 & 53\\
Northern Streamer& IC5146-W & 21:45:36 & 47:36:59 & 0.049&1.4& 3.6 & 67\\
\enddata
\tablenotetext{a}{Observation designation chosen by GBS team, denoted as Target Name in the CADC database at {\tt http://www3.cadc-ccda.hia-iha.nrc-cnrc.gc.ca/en/jcmt/}. The Proposal ID for all these observations is MJLSG36.}
\tablenotetext{b}{Central position of each observation.}
\tablenotetext{c}{Pixel-to-pixel (rms) noise within each region after mosaicking together all observations.}
\tablenotetext{d}{Effective noise per beam (i.e., point source sensitivity) within each region after mosaicking together all observations.}
\end{deluxetable}

The data reduction used for the maps presented here follows the GBS Legacy Release 1
methodology (GBS LR1) using the JCMT's Starlink software \citep{Currie14}\footnote{available
at {\tt http://starlink.eao.hawaii.edu/starlink} }, 
which is discussed by \cite{mea15}.  
The data presented here were reduced using an iterative map-making technique 
(makemap in SMURF\footnote{SMURF is a software package used for reducing JCMT observations,
and is described in more detail by \cite{jenness2013,cea13a, cea13b}}), and 
gridded to 2\arcsec\ pixels at 450\,$\mu$m and 3\arcsec\ pixels at 850\,$\mu$m.  The iterations were 
halted when the map pixels, on average, changed by $<$0.1\% of the estimated map rms. 
These initial automask reductions of each individual scan were co-added to 
form a mosaic from which a signal-to-noise ratio (SNR) mask was produced for each 
region.  The final external mask mosaic was produced from a second reduction using 
this SNR mask to define areas of emission.  In IC\,5146, the SNR mask included all pixels with signal-to-noise 
ratio of 2 or higher at 850\,$\mu$m.
Testing by our data reduction team showed similar final maps using either an 850\,$\mu$m-based
or a 450\,$\mu$m-based mask for the 450\,$\mu$m reduction, when using the SNR-based masking 
scheme described here.  Using identical masks at both wavelengths for the reduction ensures 
that the same large-scale filtering is applied to
the observations at both wavelengths (e.g., maps of the ratio of flux densities at
both wavelengths are less susceptible to differing large-scale flux recovery).
Detection of emission structure and calibration accuracy are robust within the masked regions, 
but are less certain outside of the mask \citep{mea15}.

Larger-scale structures are the most poorly recovered outside of the masked areas, while point 
sources are better recovered.  A spatial filter of 600\arcsec\ is used during both the automask and 
external mask reductions, and an additional filter of 200\arcsec\ is applied during 
the final iteration of both reductions to the areas outside of the mask.  Further testing by our data reduction 
team found that for 600\arcsec\ filtering, flux recovery is robust for sources with a Gaussian 
FWHM less than 2.5\arcmin, provided the mask is sufficiently large.  Sources between 2.5\arcmin\ and 
7.5\arcmin\ in diameter were detected, but both the flux density and the size were underestimated because 
Fourier components representing scales greater than 5\arcmin\ were removed by the filtering process.  
Detection of sources larger than 7.5\arcmin\ is dependent on the mask used for reduction.  

The data are calibrated in mJy per square arcsecond using aperture flux conversion factors (FCFs) of 
2.34\ Jy/pW/arcsec$^{2}$ $\pm 0.08$~Jy/pw/arcsec$^2$ and 
4.71\ Jy/pW/arcsec$^{2}$ $\pm 0.5$~Jy/pw/arcsec$^2$ at 850~$\mu$m and 450~$\mu$m, respectively, 
as derived from average values of JCMT calibrators \citep{dea13}. The PONG scan pattern leads to 
lower noise in the map centre and mosaic overlap regions, while data reduction and 
emission artifacts can lead to small variations in the noise over the whole map.
The pointing accuracy of the JCMT is smaller than the pixel sizes we adopt,
with current rms pointing errors of 1.2\arcsec\ in azimuth and 1.6\arcsec\ in elevation
(see http://www.eaobservatory.org/JCMT/telescope/pointing/pointing.html); JCMT pointing
accuracy in the era of SCUBA is discussed by \cite{dea08}.

The observations for IC\,5146 were taken in grade 2
($ 0.05 < \tau_{225GHz} < 0.08$) weather, corresponding to $0.21 < \tau_{850~\mu m} < 0.34$ 
\citep{dea13}, with a mean value plus standard deviation of $\tau_{225GHz}$ of 0.063 $\pm$ 0.006.
At 850\,$\mu$m, the final noise level in the mosaic is  $0.048$~mJy~arcsec$^{-2}$ per 3\arcsec\ pixel, 
corresponding to a point source sensitivity of $3.7$~mJy per 14.6\arcsec\ beam.  At 450~$\mu$m, the final noise level is 
$1.3$~mJy~arcsec$^{-2}$ per 2\arcsec\ pixel, corresponding to a point source sensitivity of  $62$~mJy per 9.8\arcsec\ beam (see Table \ref{tab:noise} for details by individual region).  The beam sizes quoted here are the effective beams determined by \cite{dea13}, and account for the fact that the beam 
shape is well-represented by the sum of a Gaussian primary beam shape and a fainter, larger Gaussian secondary beam.
 
The SCUBA-2 450\,$\mu$m and 850\,$\mu$m observations were convolved to a common beam size and
compared to estimate the temperature of the emitting dust (see Appendix A). A clump
temperature of 15\,K is adopted throughout the remainder of this paper based on these results.
Given that the CO$(J=3-2)$ rotational line lies in the middle of the 850\,$\mu$m bandpass
\citep{jea03, dea12}, an analysis was undertaken to determine the level of CO contamination for those 
limited areas of the map where CO observations also exist (see Appendix B). The contamination results 
show that none of the bright
850\,$\mu$m emission is contaminated by more than 10\%. Thus, for the remainder of this paper
we use the uncorrected 850\,$\mu$m map to determine source properties.
All of the data presented in this paper are publicly available at https://doi.org/10.11570/17.0001.

\cite{kea03} previously imaged parts of IC\,5146 at the JCMT with SCUBA \citep{hea99} at
both 450\,$\mu$m and 850\,$\mu$m, reducing the data using the SCUBA User Reduction Facility \citep[SURF;][]{jea02} and
correcting for atmospheric extinction and sky noise.  The SCUBA mapped region is $\sim14'\times2'.5$
in size and includes parts of the Northern Streamer, focusing on ridges.
The authors find several peaks of high emission (corresponding to optical extinctions of $>$ 20 mag) in their maps
that they attribute to dense prestellar structures and identify four clumps
with high optical extinctions along ridges in the region. They construct
a dust temperature map and conclude that there is a distribution of temperatures throughout the region, varying between
10\,K and 20\,K, with an average of $16.5\pm8\,$K, in agreement with the values determined in Appendix A. 
The temperatures of the cores tend toward the lower limit, $T_c \sim 12K$. 
Assuming this temperature and a dust emissivity of $\kappa_{\rm 850} = 0.01\,$cm$^2$\,g$^{-1}$, their derived core masses vary between  $4\, M_{\odot}$ and $7\, M_{\odot}$ at their adopted distance of 460\,pc (these masses increase by about a factor of four at our adopted distance of 950\,pc).
The resolution of their maps was smoothed  beyond the native JCMT resolution and so finer detail in the
cloud structure was not analysed.  These original reductions are not directly 
available for comparison. The SCUBA Legacy 
Catalog reduction \citep{dea08} of the Northern Streamer, however, is available in the archive
and was used to verify that the SCUBA-2 GBS data sets presented here are in broad agreement with
the lower sensitivity \cite{kea03} observations. 

\subsection{Extinction Map}
\label{sec:obs:ext}

\citet{c99} first published an extinction map of the IC\,5146 region 
using optical R-band star counts based on the comparison of local
stellar densities \citep{c99}. A more recent version of this IC\,5146 
extinction map (Cambr\'esy, private communication June 11, 2015) 
was derived using Two Micron All Sky Survey 
(2MASS) near-infrared H-K data to measure stellar reddening that was then used to estimate
the local extinction in the region following the technique described by 
\cite{cea02}. 
The spatial resolution of this unpublished map is 2\arcmin. Notably, known YSOs
and foreground stars have not been removed during the map construction. Nevertheless, 
the quality and resolution of this extinction map remain sufficent for the analysis required here.
Figure \ref{fig:full} shows the $A_V = 3$, 6, and 9 contours from the extinction map 
overlaid on the dust continuum SCUBA-2 850\,$\mu$m map.

\subsection{Spitzer YSOs}
\label{sec:obs:yso}

The {\it Spitzer} Space Telescope observed the IC\,5146 region \citep{hea08} using its 
InfraRed Array Camera (IRAC) and its Multiband Imaging Photometer for Spitzer 
(MIPS). The region observed by {\it Spitzer} is almost identical to the SCUBA-2 areal coverage
shown in Figure \ref{fig:full}. 
Over 200 candidate young stellar objects were identified
in the region. Those sources with both IRAC and MIPS detections have been
independently classified as Class 0+1, Flat, Class II, and Class III protostars by \cite{dea15} as part of
a larger analysis of YSOs throughout the entire Gould Belt. \cite{dea15} determine
YSO class through careful examination of the IR SEDs and their
final catalog contains an analysis of
contamination by background AGB stars, updated extinction corrections,
and revised spectral energy distributions (SEDs), improving upon previous
{\it Spitzer} YSO catalogs by \cite{hea08} and \cite{eea09}.

The {\it Spitzer} survey results combined with the 
GBS SCUBA-2 continuum data sets are shown in Figures \ref{fig:cocoon} and \ref{fig:filament}. 
There are 131 {\it Spitzer} sources within the boundaries of the SCUBA-2 observations.
Notably, the youngest YSOs, i.e., Class 0/I and Flat sources, are positioned near areas of dust emission, 
with few outliers whereas the older, Class II and III sources are more scattered.

\section{Analysis}
\label{sec:anal}

Near infrared extinction maps \citep[e.g.,][]{c99} typically trace the large-scale structure in a cloud 
complex while SCUBA-2 maps focus on denser, localized dust emission \citep[e.g.,][]{wtea16} and
the identification of sources from the {\it Spitzer} survey traces the specific
locations of YSOs \citep{dea15}. Considering together these three diverse data sets
helps us to build a better model of how star formation is influenced at
each scale.

\subsection{Large-Scale Structure}
\label{sec:anal:large}

To investigate the connection between emission in the 850\,$\mu$m SCUBA-2 
map and the observed extinction, we restrict our continuum analysis to 850\,$\mu$m pixels above a
signal to noise ratio (SNR) of 3.5, which results in a cut of pixels below a value of 0.175\,mJy\,arcsec$^{-2}$. 
This threshold is chosen to prevent the total flux density measured being 
dominated by the noise from the large number of pixels with little signal.
The extinction map, as introduced and discussed in \S\ref{sec:obs:ext}, has
a small number of zones with negative extinction caused by artifacts in the data
set. We exclude these pixels from our extinction map analysis; noting that they make up only
6\% of the total map area analyzed. 

Under the assumption that the optical characteristics of the dust grains remain the same throughout IC\,5146 and that the temperature of the dust is constant, the mass revealed by the 850\,$\mu$m map is directly proportional to the integrated flux density. Following \cite{h83}, the submillimetre-derived mass, $M_{850}$ is
\begin{equation}
M_{850}=\frac{S_{850}\ d^{2}}{\kappa_{850}\ B_{850}(T_d)},
\label{eqn:m850t}
\end{equation}
where $d$ is the distance to the cloud, $S_{850}$, $\kappa_{850}$, and $B_{850}$ are the integrated flux density, opacity, and Planck function measured at 850\,$\mu$m respectively, and $T_d$ is the dust temperature. The opacity of the dust is quite uncertain and a source of significant on-going research. Following the GBS standard 
\citep{pea15, sea15a, sea15b, rea15, kea16a, Kirk16b, mea16, rea16,Lane16}
we adopt $\kappa_{850} = 0.0125\,$cm$^{2}$\,g$^{-1}$. Taking a fiducial value for the temperature from Appendix A, $T_d = 15\,$K \citep[consistent with that used by][]{kea03}, and assuming a distance to IC\,5146, $d = 950\,$pc \citep{hea08}, Equation \ref{eqn:m850t} becomes
\begin{equation}
M_{850} = 11.3\, 
\left( {S_{850} \over {\rm Jy}} \right) 
\left( {d \over 950\, {\rm pc}} \right)^2 
\left( { \kappa_{850} \over 0.0125\, {\rm cm}^2\,{\rm g}^{-1}} \right)^{-1}
\left[ {  {\rm exp}\left( {17\,{\rm K} \over T_d} \right) - 1 \over {\rm exp}\left( {17\,{\rm K} \over 15\,{\rm K}} \right) - 1} \right]\ M_\odot.
\label{eqn:mass}
\end{equation}
Note that decreasing the fiducial dust temperature to $10\,$K would raise the derived masses by about a factor of two.

The extinction map can also be used to derive masses assuming a linear relation between extinction and column density. We adopt the \cite{sm79} ratio of $N_H / A_V = 1.87 \times 10^{21}$\,cm$^{-2}$\,mag$^{-1}$ and assume a mean molecular weight $\mu = 2.37$ \citep{Kauffmann08}.

\begin{figure}[htb]
\includegraphics[scale=0.5]{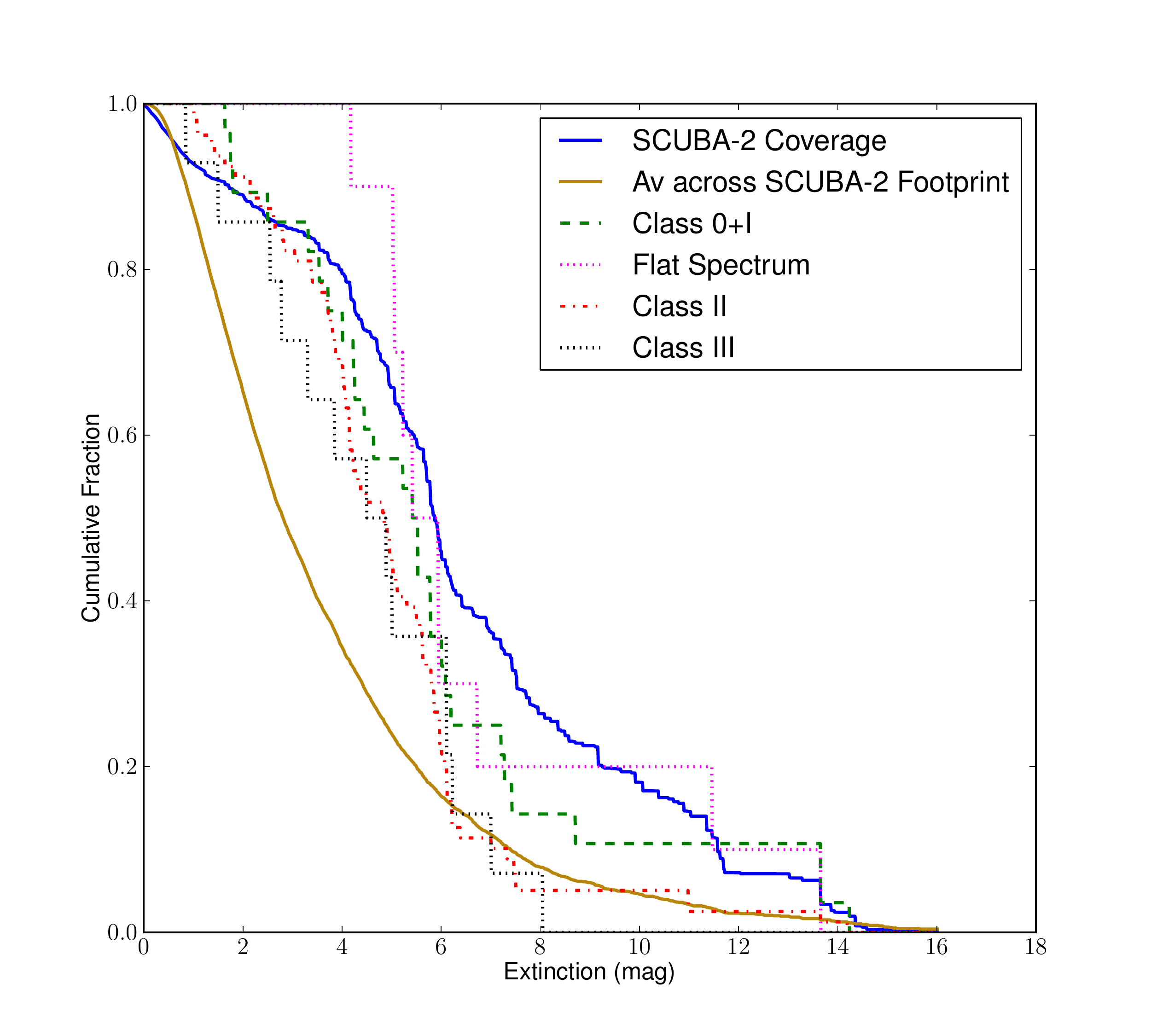}
\caption{
Cumulative mass as a function of extinction. The orange line plots
the extinction-derived mass from the Cambr\'esy extinction map restricted to the SCUBA-2 areal coverage.
The blue line plots the distribution of the submillimetre-derived mass from the 850\,$\mu$m SCUBA-2
map, as a function of the coincident Cambr\'esy extinction.
The coloured dash lines show the cumulative distributions of the YSOs, by Class, as a function of the coincident 
Cambr\'esy extinction. 
}
\label{fig:cummass}
\end{figure}

In Figure \ref{fig:cummass}, we show the cumulative fraction of mass within IC\,5146 as a function of minimum
extinction cut-off. The orange curve plots the mass derived from the extinction map using only the footprint of the 
SCUBA-2 observations. This curve reveals that most of the mass within the cloud lies at low extinction, with 50\% 
of the mass below an extinction of $A_V = 3$.  Alternatively, the blue curve plots the mass derived from the 
SCUBA-2 850\,$\mu$m map (Eqn.\ \ref{eqn:mass}), assuming that the flux density is a direct linear proxy for column 
density. The cumulative curve clearly shows that the SCUBA-2 flux predominantly traces the higher extinction regions 
within the cloud, with 85\% of the mass derived from the submillimetre continuum residing at an extinction greater than 
$A_V = 3$ and 50\% of the mass at an extinction greater than $A_V = 6$.  This result is similar to 
those found for other star-forming regions \citep[e.g.,][]{Onishi98,jdk04, Hatchell05, kea06,Enoch06,Enoch07,Konyves13}. Namely, the compact submillimetre emission
within IC\,5146 is intrinsically linked and heavily biased to regions of overall high dust column.

In total, we obtain a submillimetre-derived mass of $\sim670\,M_\odot$ for the IC\,5146 region, above the SNR$ =3.5$ cut-off. Taking only the extinction-derived mass coincident with the SCUBA-2 coverage, we obtain a mass of $16000\,M_\odot$ and a mean extinction of  $A_V = 1.7$. This latter mass is about four times the CO mass estimate derived by \citet{dea92, dea93}. The discrepancy between the CO-derived mass and the extinction-derived mass suggests one or both of the following situations: that a fraction of the extinction toward IC\,5146 is unassociated with the cloud itself, and thus the extinction mass is somewhat over-estimated, or that the CO observations were not sensitive to the extended low $A_V$ emission from IC\,5146, which is suggested by the comparison of the ${^{12}}$CO and ${^{13}}$CO images presented by \citet{dea92}. The extinction mass corresponds to roughly twenty-four times the submillimetre-derived mass.  At $A_V > 10$ we find $120\,M_\odot$ in the submillimetre map and $\sim 740\,M_\odot$ in the extinction map. Breaking down the mass estimates by sub-region within IC\,5146, we find that the extinction-derived mass coincident with the single SCUBA-2 map covering the Cocoon Nebula is $5100\,M_\odot$, or about one third of the entire mass for the IC\,5146 region.  The submillimetre-derived mass for this same region is $285\,M_\odot$, or just over 40\% of the dense gas and dust in all of IC\,5146. For the Northern Streamer, the extinction-derived mass coincident with the two SCUBA-2 PONGs  is $\sim 10^4\,M_\odot$, whereas the submillimetre-derived mass is $385\,M_\odot$. Table~\ref{tab:extinction} presents a breakdown of the extinction-derived mass, submillimetre-derived mass, and YSO count as a function of $A_V$ threshold and sub-region.

\subsection{Submillimetre Clumps}
\label{sec:anal:clumps}

We used the FellWalker algorithm \citep{b15}, part of the CUPID package \citep{bea07,berry2013} in Starlink,
to identify notable structures in the 850\,$\mu$m dust continuum map
from SCUBA-2. The FellWalker technique searches for sets of disjoint clumps, each
containing a single significant peak, using a gradient-tracing scheme. The algorithm
is qualitatively similar to the better known Clumpfind \citep{wea94} method that has
been used extensively. The Fellwalker method, however, is more stable against noise and
less susceptible to small changes in the input parameters \citep{Watson10}.

\begin{deluxetable}{cccccccccc}
\tablecolumns{10}
\tablewidth{0pc}
\tablecaption{Cumulative mass above different extinction thresholds \label{tab:extinction}}
\tablehead{
\colhead{Extinction} &
\colhead{Cloud Area} & 
\multicolumn{2}{c}{M$_{ext}$}  & 
\multicolumn{2}{c}{M$_{s2}$}  & 
\colhead{M$_{s2}$/M$_{ext}$}  & 
\colhead{C0+I+F}  &
\colhead{CII}  & 
\colhead{CIII}\\
\colhead{(mag)} &
\colhead{(\%)} &
\colhead{(M$_{\odot}$)} &
\colhead{(\%)} &
\colhead{(M$_{\odot}$)} &
\colhead{(\%)} &
\colhead{(\%)} &
\colhead{(\%)} &
\colhead{(\%)} &
\colhead{(\%)} 
}
\startdata
{\bf{S2 Coverage}}: 0 &100  &16051  &100 &670.3 &100  &4  &100  &100  &100  \\
2 &30  &10472  &65 &596.5 &89  &6  &92  &91  &86  \\
5 &6  &3844  &24 &440.7 &66  &11  &66  &46  &43  \\
10 &1  &742  &5 &121.4 &18  &16  &13  &5  &0  \\\hline
{\bf{Cocoon Neb.}}: 0 &100  &5125  &100 &285.4 &100  &6  &100  &100  &100  \\
2 &32  &3479  &68 &88.7 &31  &3  &100  &91  &89  \\
5 &5  &976  &19 &29.4 &10  &3  &79  &46  &44  \\
10 &0  &0  &0 &0.0  &0  &-- &0  &0  &0  \\\hline
{\bf{N. Streamer}}: 0 &100  &10926  &100 &384.9 &100  &4  &100  &100  &100  \\
2 &28  &6993  &64 &335.5 &87  &5  &84  &93  &80  \\
5 &6  &2868  &26 &240.0  &62  &8  &53  &43  &40  \\
10 &1  &742  &7 &121.4 &32  &16  &26  &29  &0  \\
\enddata
\end{deluxetable}

We ran FellWalker with several parameters changed from the default recommended values
to achieve two goals: to recover faint but visually distinct objects missed with the default settings
and to subdivide several larger structures that had visually apparent substructure not captured 
with the default settings.  Table\ \ref{tab:FW} lists the non-default parameters we adopted.
Note that FellWalker assumes a single global noise value for its calculations, while our
observations have some variation in noise level: the centre of each PONG is about 20\% less noisy
than the typical rms, and in the Streamer, the overlap area between the two neighbouring PONGS is
about 25\% less noisy, while the edges of the mosaic have a higher noise level than the typical rms.
This observational fact informs our two-part clump-identification
strategy: we first identify candidate clumps using FellWalker criteria that are generally more
relaxed than the default values, and then run an independent program to cull this candidate clump
list to ensure that all sources satisfy the same {\it local} SNR criteria.
To achieve our first goal of recovering faint but visually distinct objects, we
lowered the minimum flux density value allowed in clump pixels
below the default value (see the `Noise' parameter in Table~\ref{tab:FW}).  Allowing fainter
pixels to be associated with clumps
also led to the identification of some spurious noise features as clumps.  We eliminated these
false positive clumps
through the use of an automated procedure wherein each clump was required to have ten or more pixels
with a local SNR value of 3.5 or higher.  (Note that clumps passing this test are allowed to
contain additional pixels with lower local SNR values.)
This automated procedure reduced the initial FellWalker catalogue
from 273 clump candidates to 96 reliable clumps.
All of these 96 reliable clumps also passed a visual
inspection. We note that the vast majority of these clumps also have a good correspondence to the
clump catalogue obtained using the default FellWalker settings. Namely, about 10\% of the clumps in our catalogue
were either subdivided less or were not identified using the default settings.

Table \ref{tab:clumps_obs} reports the observed properties of the 96 reliable clumps, 
including the peak flux densities $F_{\rm 850}$ and $F_{\rm 450}$ in Jy\,bm$^{-1}$, 
the total integrated flux density at 850\,$\mu$m within 
the clump boundary $S_{\rm 850}$ in Jy, and the areal extent of each clump $A$ in arcsec$^2$.
As noted in \S \ref{sec:obs:S2}, the SCUBA-2  850\,$\mu$m bandpass straddles the CO($J=3-2$) rotational line 
that can result in
CO contamination of the continuum flux. Where possible, the peak and total flux density at 850\,$\mu$m associated with each clump
have been investigated for possible CO contamination (Appendix B) and in all cases the contamination values
are found to be less than 10\%. The results shown are not corrected for CO contamination.

Table \ref{tab:850der} presents the derived properties of the clumps.
The effective radius of each clump,
$R_C$ in pc, is derived through equating the area within the clump boundary with that of a circular aperture.
The masses of the clumps are computed using Equation \ref{eqn:mass}
under the assumption that $T_d = 15\,$K (\S \ref{sec:obs:S2} and Appendix A).
Decreasing the dust temperature to $10\,$K raises the derived masses by about
a factor of two.
The masses span a range from 0.5\,$M_\odot$ to 116\,$M_\odot$,
the mean clump mass is  8\,$M_\odot$, and the median clump mass is 3.7\,$M_\odot$. The total mass in clumps is
$\sim 750\,M_\odot$, slightly larger than the submillimetre-derived mass found in the previous section.
This difference is because we allow the clump boundaries to include not only the bright central
emission, but also extended more diffuse emission (below a local SNR=3.5 level) that is clearly associated.
In our analysis in Section~4.1, we use a conservative global SNR=3.5 threshold to prevent noise spikes
at lower SNR levels from being included in our results.  Since most of the area of our map appears to
lack real emission, noise spikes would make a significant contribution to flux measured
at levels below an SNR of 3.5.

\begin{deluxetable}{llc}
\tablecolumns{3}
\tablewidth{0pc}
\tablecaption{FellWalker Parameters \label{tab:FW}}

\tablehead{
\colhead{Name}&
\colhead{Descriptor}&
\colhead{Value}
}
\startdata
AllowEdge & exclude clumps touching the noisier map edge&0.00\\
CleanIter & smooth clump edges & 15.0\\
RMS& as measured in the map & 0.05\\
FlatSlope & increase the gradient required for associating pixels with a peak & 2.0{*}RMS\\
MinDip & reduce the difference between peaks required to have  multiple objects& 1.5{*}RMS\\
MaxJump & reduce the search radius to combine peaks into a single object &2.0\\
MinPix & minimum number of pixels per clump& 10.0\\
Noise & allow fainter pixels to be associated with a peak &0.35{*}RMS\\
\enddata
\end{deluxetable}

\begin{deluxetable}{cccccccr}
\tablecolumns{8}
\tablewidth{0pt}
\tabletypesize{\scriptsize}
\tablecaption{Identified 850\,$\mu$m Clumps \label{tab:clumps_obs}}
\tablehead{
\colhead{Clump\tablenotemark{a}}&
\colhead{Source Name\tablenotemark{b}}&
\colhead{R.A.\tablenotemark{c}}&
\colhead{Dec.\tablenotemark{c}}&
\colhead{Peak Flux\tablenotemark{d}}&
\colhead{Peak Flux\tablenotemark{d}}&
\colhead{Total Flux\tablenotemark{d}}&
\colhead{Area\tablenotemark{d}}\\
\colhead{Number}&
\colhead{(MJLSG...)}&
\colhead{(J2000.0)} &
\colhead{(J2000.0)} &
\colhead{$F_{850}$ (Jy/Bm)} &
\colhead{$F_{450}$ (Jy/Bm)} &
\colhead{$S_{850}$ (Jy)} &
\colhead{$A$ (arcsec$^{2}$)}
}
\startdata
1 & J214428.6+473616 & 21:44:28.57 & 47:36:15.80 & 0.05 & 0.46 & 0.16 & 2457\\
2 & J214439.4+474416 & 21:44:39.35 & 47:44:16.02 & 0.07 & 0.47 & 0.14 & 2151\\
3 & J214439.4+473522 & 21:44:39.40 & 47:35:22.03 & 0.05 & 0.48 & 0.09 & 1305\\
4 & J214442.8+474641 & 21:44:42.80 & 47:46:41.03 & 0.36 & 1.54 & 0.92 & 5292\\
5 & J214443.6+474532 & 21:44:43.59 & 47:45:32.27 & 0.08 & 0.52 & 0.32 & 3051\\
6 & J214448.1+474458 & 21:44:48.15 & 47:44:57.60 & 0.09 & 0.49 & 0.31 & 3573\\
7 & J214448.9+473649 & 21:44:48.93 & 47:36:48.82 & 0.07 & 0.46 & 0.27 & 3168\\
8 & J214452.2+474032 & 21:44:52.16 & 47:40:31.76 & 0.41 & 1.62 & 2.10 & 8163\\
9 & J214454.3+474214 & 21:44:54.25 & 47:42:14.36 & 0.06 & 0.40 & 0.08 & 990\\
10 & J214454.8+473903 & 21:44:54.79 & 47:39:02.51 & 0.08 & 0.45 & 0.07 & 693\\
11 & J214455.6+473927 & 21:44:55.61 & 47:39:26.75 & 0.16 & 0.65 & 0.42 & 1836\\
12 & J214458.2+474154 & 21:44:58.17 & 47:41:54.48 & 0.19 & 0.67 & 0.56 & 3258\\
13 & J214458.2+474003 & 21:44:58.18 & 47:40:03.48 & 0.16 & 0.69 & 1.44 & 5832\\
14 & J214458.8+473434 & 21:44:58.80 & 47:34:33.65 & 0.08 & 0.51 & 0.16 & 1566\\
15 & J214459.7+473937 & 21:44:59.74 & 47:39:36.92 & 0.10 & 0.52 & 0.15 & 1332\\
16 & J214460.7+473355 & 21:44:60.68 & 47:33:55.18 & 0.08 & 0.69 & 0.29 & 2340\\
17 & J214462.9+473302 & 21:44:62.90 & 47:33:01.81 & 0.12 & 0.64 & 0.37 & 2574\\
18 & J214506.5+474357 & 21:45:06.45 & 47:43:56.82 & 0.10 & 0.49 & 0.51 & 5184\\
19 & J214507.5+473345 & 21:45:07.53 & 47:33:45.11 & 0.08 & 0.54 & 0.20 & 1719\\
20 & J214507.7+473809 & 21:45:07.70 & 47:38:09.17 & 0.08 & 0.52 & 0.24 & 2493\\
21 & J214508.2+473306 & 21:45:08.22 & 47:33:06.31 & 0.43 & 2.01 & 1.25 & 5562\\
22 & J214512.9+473723 & 21:45:12.87 & 47:37:22.60 & 0.07 & 0.50 & 0.27 & 2601\\
23 & J214513.3+473235 & 21:45:13.34 & 47:32:34.73 & 0.09 & 0.55 & 0.34 & 3033\\
24 & J214514.1+473705 & 21:45:14.10 & 47:37:04.95 & 0.07 & 0.46 & 0.10 & 1368\\
25 & J214527.7+473545 & 21:45:27.66 & 47:35:44.64 & 0.11 & 0.50 & 0.28 & 2241\\
26 & J214531.1+473616 & 21:45:31.14 & 47:36:15.57 & 0.10 & 0.61 & 0.52 & 3636\\
27 & J214535.4+473541 & 21:45:35.38 & 47:35:40.70 & 0.11 & 0.67 & 0.60 & 3789\\
28 & J214550.8+473736 & 21:45:50.81 & 47:37:35.72 & 0.06 & 0.43 & 0.16 & 2061\\
29 & J214552.3+473745 & 21:45:52.27 & 47:37:45.10 & 0.07 & 0.45 & 0.11 & 1512\\
30 & J214558.8+473602 & 21:45:58.75 & 47:36:01.74 & 0.17 & 0.74 & 0.98 & 5940\\
31 & J214561.5+473550 & 21:45:61.45 & 47:35:50.42 & 0.12 & 0.45 & 0.32 & 2511\\
32 & J214617.8+473345 & 21:46:17.75 & 47:33:45.43 & 0.07 & 0.53 & 0.12 & 1305\\
33 & J214620.4+473407 & 21:46:20.37 & 47:34:07.06 & 0.08 & 0.61 & 0.34 & 3771\\
34 & J214632.8+473355 & 21:46:32.84 & 47:33:55.02 & 0.06 & 0.65 & 0.26 & 3591\\
35 & J214649.8+473314 & 21:46:49.82 & 47:33:13.90 & 0.07 & 0.46 & 0.38 & 4311\\
36 & J214650.0+473429 & 21:46:49.96 & 47:34:28.93 & 0.08 & 0.41 & 0.45 & 4536\\
37 & J214651.3+473953 & 21:46:51.33 & 47:39:53.24 & 0.07 & 0.44 & 0.33 & 4500\\
38 & J214656.9+473312 & 21:46:56.94 & 47:33:12.48 & 0.06 & 0.38 & 0.28 & 3438\\
39 & J214660.2+473949 & 21:46:60.24 & 47:39:49.21 & 0.07 & 0.33 & 0.11 & 1206\\
40 & J214704.4+473241 & 21:47:04.41 & 47:32:41.12 & 0.10 & 0.45 & 0.18 & 1125\\
41 & J214706.2+473936 & 21:47:06.21 & 47:39:35.50 & 0.12 & 0.45 & 0.77 & 5472\\
42 & J214706.4+473306 & 21:47:06.43 & 47:33:05.55 & 0.14 & 0.51 & 1.35 & 10863\\
43 & J214708.9+473157 & 21:47:08.95 & 47:31:57.09 & 0.06 & 0.37 & 0.24 & 2592\\
44 & J214709.2+473933 & 21:47:09.19 & 47:39:33.14 & 0.09 & 0.44 & 0.62 & 5139\\
45 & J214720.5+473433 & 21:47:20.48 & 47:34:32.52 & 0.05 & 0.39 & 0.09 & 1152\\
46 & J214722.7+473342 & 21:47:22.66 & 47:33:41.97 & 0.16 & 0.60 & 0.78 & 3996\\
47 & J214722.8+473212 & 21:47:22.84 & 47:32:12.01 & 0.70 & 2.92 & 10.34 & 21312\\
48 & J214724.8+473103 & 21:47:24.75 & 47:31:03.40 & 0.19 & 0.63 & 0.72 & 3204\\
49 & J214724.8+473815 & 21:47:24.78 & 47:38:15.41 & 0.07 & 0.40 & 0.26 & 3105\\
50 & J214726.6+473534 & 21:47:26.58 & 47:35:33.78 & 0.06 & 0.31 & 0.12 & 1566\\
51 & J214729.0+473734 & 21:47:29.01 & 47:37:34.28 & 0.06 & 0.39 & 0.18 & 2304\\
52 & J214730.3+473202 & 21:47:30.26 & 47:32:01.53 & 0.19 & 0.67 & 0.92 & 4275\\
53 & J214730.6+472935 & 21:47:30.55 & 47:29:34.59 & 0.07 & 0.38 & 0.21 & 2556\\
54 & J214733.5+473944 & 21:47:33.51 & 47:39:44.18 & 0.05 & 0.42 & 0.05 & 837\\
55 & J214750.6+473751 & 21:47:50.64 & 47:37:50.54 & 0.08 & 0.44 & 0.28 & 3024\\
56 & J214753.9+473736 & 21:47:53.93 & 47:37:36.16 & 0.07 & 0.50 & 0.33 & 2880\\
57 & J214759.1+473625 & 21:47:59.11 & 47:36:25.14 & 0.29 & 1.48 & 1.11 & 6426\\
58 & J214812.8+473625 & 21:48:12.75 & 47:36:24.63 & 0.07 & 0.42 & 0.24 & 2862\\
59 & J214842.8+472560 & 21:48:42.81 & 47:25:59.77 & 0.07 & 0.35 & 0.08 & 1044\\
60 & J214846.1+472551 & 21:48:46.07 & 47:25:51.30 & 0.08 & 0.44 & 0.38 & 3150\\
61 & J214846.1+472515 & 21:48:46.13 & 47:25:15.31 & 0.10 & 0.47 & 0.37 & 2943\\
62 & J214852.0+473134 & 21:48:52.05 & 47:31:34.26 & 0.11 & 0.50 & 0.64 & 5553\\
63 & J214854.8+472723 & 21:48:54.80 & 47:27:22.69 & 0.05 & 0.39 & 0.11 & 1647\\
64 & J214856.3+473026 & 21:48:56.30 & 47:30:25.93 & 0.17 & 0.68 & 0.72 & 5112\\
65 & J214923.5+472406 & 21:49:23.47 & 47:24:05.95 & 0.05 & 0.42 & 0.05 & 666\\
66 & J214924.3+472851 & 21:49:24.26 & 47:28:51.06 & 0.27 & 0.94 & 1.43 & 7146\\
67 & J214931.0+472607 & 21:49:30.99 & 47:26:06.99 & 0.07 & 0.49 & 0.25 & 2979\\
68 & J214931.5+472725 & 21:49:31.48 & 47:27:25.06 & 0.14 & 0.66 & 0.74 & 3996\\
69 & J214933.9+472710 & 21:49:33.86 & 47:27:10.39 & 0.18 & 0.71 & 0.65 & 3114\\
70 & J214936.8+472711 & 21:49:36.82 & 47:27:10.78 & 0.10 & 0.49 & 0.27 & 1971\\
71 & J215235.8+471314 & 21:52:35.75 & 47:13:13.99 & 0.05 & 0.47 & 0.12 & 1638\\
72 & J215238.4+471438 & 21:52:38.38 & 47:14:38.08 & 0.18 & 0.96 & 0.58 & 4833\\
73 & J215307.0+471521 & 21:53:06.95 & 47:15:20.85 & 0.17 & 0.67 & 1.76 & 7785\\
74 & J215307.3+471424 & 21:53:07.25 & 47:14:23.86 & 0.12 & 0.65 & 0.39 & 2367\\
75 & J215312.6+471457 & 21:53:12.55 & 47:14:56.95 & 0.21 & 0.84 & 2.38 & 8775\\
76 & J215312.9+471136 & 21:53:12.87 & 47:11:35.96 & 0.06 & 0.45 & 0.17 & 2655\\
77 & J215314.6+471621 & 21:53:14.60 & 47:16:20.99 & 0.18 & 0.73 & 2.05 & 9630\\
78 & J215315.5+471821 & 21:53:15.46 & 47:18:21.00 & 0.08 & 0.54 & 0.39 & 3654\\
79 & J215317.8+471939 & 21:53:17.81 & 47:19:39.03 & 0.11 & 0.65 & 1.17 & 7893\\
80 & J215324.6+471730 & 21:53:24.61 & 47:17:30.11 & 0.08 & 0.50 & 0.24 & 2466\\
81 & J215331.4+472203 & 21:53:31.38 & 47:22:03.17 & 0.17 & 0.91 & 1.59 & 7974\\
82 & J215333.2+471415 & 21:53:33.17 & 47:14:15.18 & 0.24 & 0.86 & 1.46 & 6156\\
83 & J215334.6+472057 & 21:53:34.63 & 47:20:57.18 & 0.18 & 0.80 & 2.66 & 9783\\
84 & J215335.5+471727 & 21:53:35.52 & 47:17:27.19 & 0.16 & 0.67 & 0.55 & 2088\\
85 & J215336.4+471506 & 21:53:36.41 & 47:15:06.19 & 0.07 & 0.45 & 0.23 & 1908\\
86 & J215336.7+471903 & 21:53:36.70 & 47:19:03.19 & 0.32 & 1.29 & 1.33 & 5175\\
87 & J215336.7+471415 & 21:53:36.71 & 47:14:15.19 & 0.10 & 0.71 & 0.30 & 2358\\
88 & J215337.3+471736 & 21:53:37.29 & 47:17:36.19 & 0.14 & 0.63 & 1.12 & 4509\\
89 & J215337.6+471948 & 21:53:37.58 & 47:19:48.19 & 0.31 & 1.16 & 1.82 & 6129\\
90 & J215340.8+471527 & 21:53:40.83 & 47:15:27.20 & 0.05 & 0.43 & 0.17 & 2259\\
91 & J215345.3+471357 & 21:53:45.25 & 47:13:57.20 & 0.05 & 0.50 & 0.10 & 1485\\
92 & J215345.8+471736 & 21:53:45.84 & 47:17:36.20 & 0.10 & 0.60 & 0.49 & 3357\\
93 & J215350.5+471233 & 21:53:50.55 & 47:12:33.18 & 0.10 & 0.69 & 0.91 & 6192\\
94 & J215350.5+471348 & 21:53:50.55 & 47:13:48.18 & 0.21 & 0.94 & 1.74 & 5652\\
95 & J215354.7+471321 & 21:53:54.67 & 47:13:21.16 & 0.21 & 0.95 & 1.51 & 5445\\
96 & J215355.9+471354 & 21:53:55.85 & 47:13:54.15 & 0.20 & 0.91 & 1.39 & 6354\\
\enddata
\tablenotetext{a}{Clump observation designation}
\tablenotetext{b}{The source name is based on the coordinates of the peak emission location of each object in right ascension and declination: Jhhmmss.s$\pm$ddmmss.}
\tablenotetext{c}{Peak position, at 850\,$\mu$m, for each clump.}
\tablenotetext{d}{Peak flux at 850~$\mu$m and 450~$\mu$m within the clump boundary, total flux
	at 850~$\mu$m within the clump boundary, and the area spanned by the clump.}
\end{deluxetable}

\begin{deluxetable}{crrcccccc}
\tablecolumns{9}
\tablewidth{0pt}
\tabletypesize{\scriptsize}
\tablecaption{Derived Properties of the Identified 850\,$\mu$m Clumps\label{tab:850der}}
\tablehead{
\colhead{Clump\tablenotemark{a}}&
\colhead{Total Mass\tablenotemark{b}}&
\colhead{Jeans Mass\tablenotemark{b}}&
\colhead{Clump Radius\tablenotemark{b}}&
\colhead{Concentration\tablenotemark{b}}&
\colhead{Total YSOs}&
\colhead{Class0+I/Flat}&
\colhead{Class II/III}&
\colhead{Class0+I/Flat}
\\
\colhead{Number}&
\colhead{$M_{850}$ ($M_{\odot}$)}&
\colhead{$M_J$ ($M_{\odot}$)} &
\colhead{$R_c$ (parsec)} &
\colhead{}&
\colhead{Contained\tablenotemark{c}}& 
\colhead{Contained\tablenotemark{c}}&
\colhead{near Peak\tablenotemark{c}}&
\colhead{near Peak\tablenotemark{c}}
}
\startdata
1 & 1.77 & 5.24 & 0.13 & 0.71 & 0 & 0 & 0 & 0\\
2 & 1.60 & 4.91 & 0.12 & 0.78 & 0 & 0 & 0 & 0\\
3 & 0.98 & 3.82 & 0.09 & 0.68 & 0 & 0 & 0 & 0\\
4 & 10.40 & 7.70 & 0.19 & 0.88 & 2 & 1 & 0 & 1\\
5 & 3.65 & 5.84 & 0.14 & 0.69 & 0 & 0 & 0 & 0\\
6 & 3.50 & 6.32 & 0.16 & 0.76 & 1 & 1 & 0 & 1\\
7 & 3.08 & 5.95 & 0.15 & 0.70 & 0 & 0 & 0 & 0\\
8 & 23.67 & 9.56 & 0.23 & 0.85 & 3 & 2 & 0 & 2\\
9 & 0.95 & 3.33 & 0.08 & 0.64 & 0 & 0 & 0 & 0\\
10 & 0.75 & 2.79 & 0.07 & 0.72 & 0 & 0 & 0 & 0\\
11 & 4.71 & 4.53 & 0.11 & 0.66 & 0 & 0 & 0 & 0\\
12 & 6.26 & 6.04 & 0.15 & 0.78 & 1 & 1 & 0 & 1\\
13 & 16.21 & 8.08 & 0.20 & 0.63 & 0 & 0 & 0 & 0\\
14 & 1.77 & 4.19 & 0.10 & 0.71 & 0 & 0 & 0 & 0\\
15 & 1.71 & 3.86 & 0.09 & 0.74 & 0 & 0 & 0 & 0\\
16 & 3.25 & 5.12 & 0.13 & 0.63 & 0 & 0 & 0 & 0\\
17 & 4.17 & 5.37 & 0.13 & 0.72 & 1 & 1 & 0 & 1\\
18 & 5.69 & 7.62 & 0.19 & 0.76 & 0 & 0 & 0 & 0\\
19 & 2.27 & 4.39 & 0.11 & 0.65 & 0 & 0 & 0 & 0\\
20 & 2.73 & 5.28 & 0.13 & 0.69 & 0 & 0 & 0 & 0\\
21 & 14.10 & 7.89 & 0.19 & 0.88 & 1 & 1 & 0 & 1\\
22 & 3.00 & 5.40 & 0.13 & 0.64 & 0 & 0 & 0 & 0\\
23 & 3.85 & 5.83 & 0.14 & 0.70 & 0 & 0 & 0 & 0\\
24 & 1.15 & 3.91 & 0.10 & 0.73 & 0 & 0 & 0 & 0\\
25 & 3.14 & 5.01 & 0.12 & 0.72 & 0 & 0 & 0 & 0\\
26 & 5.79 & 6.38 & 0.16 & 0.67 & 1 & 1 & 0 & 1\\
27 & 6.72 & 6.51 & 0.16 & 0.67 & 0 & 0 & 0 & 0\\
28 & 1.81 & 4.80 & 0.12 & 0.69 & 0 & 0 & 0 & 0\\
29 & 1.21 & 4.11 & 0.10 & 0.74 & 0 & 0 & 0 & 0\\
30 & 11.05 & 8.15 & 0.20 & 0.76 & 1 & 1 & 0 & 1\\
31 & 3.64 & 5.30 & 0.13 & 0.75 & 0 & 0 & 0 & 0\\
32 & 1.38 & 3.82 & 0.09 & 0.67 & 0 & 0 & 0 & 0\\
33 & 3.86 & 6.50 & 0.16 & 0.72 & 0 & 0 & 0 & 0\\
34 & 2.92 & 6.34 & 0.16 & 0.72 & 0 & 0 & 0 & 0\\
35 & 4.28 & 6.95 & 0.17 & 0.70 & 0 & 0 & 0 & 0\\
36 & 5.11 & 7.13 & 0.18 & 0.71 & 0 & 0 & 0 & 0\\
37 & 3.68 & 7.10 & 0.17 & 0.76 & 0 & 0 & 0 & 0\\
38 & 3.21 & 6.20 & 0.15 & 0.69 & 0 & 0 & 0 & 0\\
39 & 1.24 & 3.67 & 0.09 & 0.70 & 0 & 0 & 0 & 0\\
40 & 2.01 & 3.55 & 0.09 & 0.64 & 0 & 0 & 0 & 0\\
41 & 8.67 & 7.83 & 0.19 & 0.71 & 1 & 1 & 0 & 1\\
42 & 15.20 & 11.03 & 0.27 & 0.78 & 1 & 1 & 0 & 0\\
43 & 2.72 & 5.39 & 0.13 & 0.64 & 1 & 0 & 0 & 0\\
44 & 6.96 & 7.58 & 0.19 & 0.67 & 0 & 0 & 0 & 0\\
45 & 1.05 & 3.59 & 0.09 & 0.61 & 0 & 0 & 0 & 0\\
46 & 8.78 & 6.69 & 0.16 & 0.71 & 0 & 0 & 0 & 0\\
47 & 116.36 & 15.44 & 0.38 & 0.83 & 5 & 4 & 0 & 2\\
48 & 8.15 & 5.99 & 0.15 & 0.71 & 0 & 0 & 0 & 0\\
49 & 2.94 & 5.90 & 0.14 & 0.69 & 0 & 0 & 0 & 0\\
50 & 1.32 & 4.19 & 0.10 & 0.71 & 0 & 0 & 0 & 0\\
51 & 1.99 & 5.08 & 0.12 & 0.71 & 0 & 0 & 0 & 0\\
52 & 10.31 & 6.92 & 0.17 & 0.73 & 2 & 0 & 0 & 0\\
53 & 2.35 & 5.35 & 0.13 & 0.70 & 0 & 0 & 0 & 0\\
54 & 0.61 & 3.06 & 0.08 & 0.71 & 0 & 0 & 0 & 0\\
55 & 3.13 & 5.82 & 0.14 & 0.73 & 0 & 0 & 0 & 0\\
56 & 3.69 & 5.68 & 0.14 & 0.63 & 1 & 1 & 0 & 0\\
57 & 12.51 & 8.48 & 0.21 & 0.85 & 1 & 1 & 0 & 1\\
58 & 2.66 & 5.66 & 0.14 & 0.71 & 0 & 0 & 0 & 0\\
59 & 0.88 & 3.42 & 0.08 & 0.74 & 0 & 0 & 0 & 0\\
60 & 4.29 & 5.94 & 0.15 & 0.65 & 0 & 0 & 0 & 0\\
61 & 4.16 & 5.74 & 0.14 & 0.71 & 0 & 0 & 0 & 0\\
62 & 7.23 & 7.88 & 0.19 & 0.75 & 0 & 0 & 0 & 0\\
63 & 1.19 & 4.29 & 0.11 & 0.70 & 0 & 0 & 0 & 0\\
64 & 8.11 & 7.56 & 0.19 & 0.80 & 0 & 0 & 0 & 0\\
65 & 0.53 & 2.73 & 0.07 & 0.68 & 0 & 0 & 0 & 0\\
66 & 16.13 & 8.94 & 0.22 & 0.82 & 0 & 0 & 0 & 0\\
67 & 2.84 & 5.77 & 0.14 & 0.73 & 0 & 0 & 0 & 0\\
68 & 8.36 & 6.69 & 0.16 & 0.69 & 0 & 0 & 0 & 0\\
69 & 7.36 & 5.90 & 0.15 & 0.72 & 0 & 0 & 0 & 0\\
70 & 3.06 & 4.70 & 0.12 & 0.68 & 0 & 0 & 0 & 0\\
71 & 1.33 & 4.28 & 0.11 & 0.65 & 0 & 0 & 0 & 0\\
72 & 6.52 & 7.35 & 0.18 & 0.84 & 2 & 1 & 0 & 1\\
73 & 19.82 & 9.33 & 0.23 & 0.67 & 0 & 0 & 0 & 0\\
74 & 4.42 & 5.15 & 0.13 & 0.68 & 1 & 1 & 0 & 1\\
75 & 26.83 & 9.91 & 0.24 & 0.69 & 1 & 0 & 0 & 0\\
76 & 1.87 & 5.45 & 0.13 & 0.72 & 0 & 0 & 0 & 0\\
77 & 23.10 & 10.38 & 0.25 & 0.71 & 0 & 0 & 0 & 0\\
78 & 4.42 & 6.40 & 0.16 & 0.66 & 0 & 0 & 0 & 0\\
79 & 13.22 & 9.40 & 0.23 & 0.68 & 0 & 0 & 0 & 0\\
80 & 2.74 & 5.25 & 0.13 & 0.71 & 0 & 0 & 0 & 0\\
81 & 17.85 & 9.45 & 0.23 & 0.71 & 1 & 1 & 0 & 1\\
82 & 16.44 & 8.30 & 0.20 & 0.76 & 3 & 2 & 0 & 1\\
83 & 29.97 & 10.46 & 0.26 & 0.63 & 1 & 1 & 0 & 1\\
84 & 6.24 & 4.83 & 0.12 & 0.60 & 1 & 1 & 0 & 0\\
85 & 2.64 & 4.62 & 0.11 & 0.58 & 1 & 0 & 0 & 0\\
86 & 14.99 & 7.61 & 0.19 & 0.81 & 2 & 2 & 0 & 1\\
87 & 3.35 & 5.14 & 0.13 & 0.69 & 0 & 0 & 0 & 0\\
88 & 12.65 & 7.10 & 0.17 & 0.58 & 0 & 0 & 0 & 0\\
89 & 20.54 & 8.28 & 0.20 & 0.77 & 1 & 1 & 0 & 0\\
90 & 1.88 & 5.03 & 0.12 & 0.67 & 2 & 0 & 1 & 0\\
91 & 1.17 & 4.08 & 0.10 & 0.66 & 0 & 0 & 0 & 0\\
92 & 5.54 & 6.13 & 0.15 & 0.65 & 0 & 0 & 0 & 0\\
93 & 10.25 & 8.33 & 0.20 & 0.64 & 1 & 0 & 0 & 0\\
94 & 19.54 & 7.95 & 0.20 & 0.65 & 1 & 1 & 0 & 1\\
95 & 17.01 & 7.81 & 0.19 & 0.68 & 0 & 0 & 0 & 0\\
96 & 15.66 & 8.43 & 0.21 & 0.74 & 2 & 1 & 0 & 0\\
\enddata
\tablenotetext{a}{Clump observation designation.}
\tablenotetext{b}{See text for definitions.}
\tablenotetext{c}{``Contained" refers to YSOs that reside within the clump boundary. `Near peak'' refers to those YSOs that reside within 15$\arcsec$ of the clump's peak position.}
\end{deluxetable}

Figures \ref{fig:cocoon} and \ref{fig:filament} use contours to show the clump boundaries
within the two IC\,5146 molecular cloud regions. Within the Cocoon Nebula, the majority of the clumps merge 
together to create a broken ring around the central star cluster while in the Northern Streamer, the 
clumps fan out along the known filaments uncovered by {\it Herschel} \citep{aea11}. In both regions, however,
the distribution of clumps typically generate one-dimensional sequences and relatively straight filamentary chains.

Although the single most massive clump lies in the Northern Streamer, the typical clump in the Cocoon Nebula 
is about twice as massive (mean 11.5\,$M_\odot$, median 6.6\,$M_\odot$) as that found in the Northern Streamer 
(mean 6.5\,$M_\odot$, median 3.2\,$M_\odot$), assuming that the temperatures and
dust properties are the same across all of IC\,5146. Many of the clumps in IC\,5146 are closely related to the YSOs,
especially the Class 0 and I sources (see \S\ \ref{sec:anal:yso}).  Of the 70 clumps in the Northern 
Streamer (Clumps 1 to 70 in Table \ref{tab:clumps_obs}), fifteen (21\,\%) have at least one associated YSO.  
In contrast, of the 26 clumps observed in the Cocoon Nebula (Clumps 71 to 96), fourteen
(54\,\%) harbour at least one YSO within the boundaries. At first glance, this suggests that star formation is
more active in the Cocoon Nebula. It is also possible, however, that the earliest stage of star (clump) formation 
is ramping down in the Cocoon Nebula and therefore the majority of the remaining clumps are
presently star-forming. In the Northern Streamer, a smaller fraction of clumps host embedded YSOs and
almost all of the YSOs in the region are still heavily embedded in the dust continuum, implying
an earlier evolutionary time (see also \S \ref{sec:anal:yso}). 


Although these clumps are likely to have additional non-thermal support, given their large size 
and mass, it is interesting to compare them against known static isothermal models, such as 
Bonnor-Ebert (BE) spheres \citep{e55,b56}. BE sphere models denote a continuum of solutions for equilibrium 
self-gravitating isothermal spheres with external bounding pressure, from very low mass objects that
have an almost constant density throughout to critical models that are on the very edge of gravitational 
collapse and have a large variation in density between the centre and the edge. This continuum of 
models can be represented by a single observational measure, the concentration parameter, as 
described by \cite{jea00} and \cite{kea06}:
\begin{equation}
C = 1 - \left({ 1.13\,B^2\, \over A}\right) \left({S_{850} \over  F_{850}}\right),
\label{eqn:c}
\end{equation}
where $B= 14.6\arcsec$ is the effective 850 $\mu m$ JCMT beamsize \citep{dea13}. To be stable against collapse, 
isothermal clumps with concentrations above 0.72 require additional support mechanisms such as pressure from 
magnetic fields. The concentration for each of the clumps in IC\,5146  is included in Table \ref{tab:850der} .
The concentrations range from 0.58 to 0.88, with a mean value of 0.70. Thus, the typical submillimetre clump 
appears on the verge of gravo-thermal instability based on concentration. It is worth noting that this analysis does
not depend on the inferred temperature or dust emissivity  (see Eqn.\ \ref{eqn:c}) but does 
require that these properties remain constant throughout the clump. Furthermore, the concentration parameter is
sensitive to the derived radius of the clump and thus choosing lower surface brightness thresholds for
clump boundaries is likely to result in larger derived concentrations. As a result, individual concentration values should
be treated with caution while the variations in concentration between clumps and across ensembles of clumps provides
indications of the varying importance of gravity and non-thermal properties.

For all the clumps in IC\,5146 we also compare the derived mass, using Equation \ref{eqn:mass}, against the
maximum (Jeans-critical) mass of a stable thermally-supported sphere of the same size $R_c$ following the
strategy of \cite{sea10} and assuming a gas temperature of 15\,K consistent with that determined from the dust
\begin{equation}
M_J =  29 \left( { T\over 15\,{\rm K}} \right) \left( {R_c \over 0.07\,{\rm pc}} \right) M_\odot.
\end{equation}
As shown in Table \ref{tab:850der}, the range of Jeans masses, $2.7< M_J/M_\odot < 15.4$, is significantly smaller than the
range of clump masses $M_{850}$. Nevertheless,  a majority of the observed clumps appear to be Jeans stable according 
to this criterion: i.e., 62 of 96 or 65\% have $M_{850}/M_J < 1$, with a noticeable difference between the two regions. In the Northern 
Streamer, 51 of 70 or 73\% are stable by this criterion whereas in the Cocoon Nebula only 11 of 26 or 42\% appear stable.
Furthermore, the subset of 29 clumps across IC\,5146 harbouring embedded protostars show a propensity for instability, with 20 of 29 or 
69\% having $M_{850}/M_J > 1$. The Jeans stability argument is extremely 
sensitive to the assumed temperature given that the Jeans mass increases and the dust continuum mass decreases with 
increasing temperature.  Nevertheless, assuming that the properties of clumps are similar across IC\,5146, the Jeans stability ratio ($M_{850}/M_J$) allows for an ordering of the importance of self-gravity within clumps, with the highest ratios denoting the most gravitationally unstable clumps.

\subsection{Dust Continuum and YSOs}
\label{sec:anal:yso}

{\it Spitzer} observed the IC\,5146 region \citep{hea08} and identified 131 YSOs within the SCUBA-2 areal coverage, classified
as either Class 0/I (28), Flat (10), Class II (79), or Class III (14) \citep{dea15}.  Of these, 38 reside
in the Northern Streamer, with 19 Class 0/I and Flat protostars versus 19 Class II+III more evolved YSOs.
The Cocoon Nebula has 91 YSOs in total, divided into 19 Class 0/I and Flat protostars versus 74 Class II+IIIs. 
Tables \ref{tab:protos} and \ref{tab:disks} detail the location, the underlying 450\,$\mu$m and 850\,$\mu$m flux density, and
the associated near infrared extinction for {\it Spitzer} ``protostars" (Class 0/I and Flat) and ``evolved sources" (Class II+III), respectively. 
These tables also note the identifier of the clump in which each YSO resides 
as well as both the offset distance to the nearest clump peak and that clump-identifier. 
Even with the greater number of YSOs being in the Cocoon Nebula, only twenty (21\,\%) of the
YSOs are associated with clumps identified in the region (see also Figure \ref{fig:cocoon}). 
Broken down by type, twelve of the protostars (63\%) and eight of the more-evolved YSOs (11\%) are coincident with
clumps. Alternatively, looking at the Northern Streamer, twenty-three YSOs (60\,\%) are still within
the observed clumps. This number can be broken down to seventeen protostars (89\%) and six more-evolved sources
(32\%). Hence, independent of which evolutionary stage of YSO one compares the Northern Streamer appears 
to be less evolved than the Cocoon Nebula (see also \S \ref{sec:anal:clumps}).

\begin{deluxetable}{ccccrrrrrr}
\tablecolumns{10}
\tablewidth{0pc}
\tabletypesize{\scriptsize}
\tablecaption{{\it Spitzer} Protostars: Flux and Extinction\label{tab:protos}}
\tablehead{
\colhead{YSO Index\tablenotemark{a}}&
\colhead{R.A.\tablenotemark{b}}&
\colhead{Dec.\tablenotemark{b}}&
\colhead{Class Type\tablenotemark{c}}&
\colhead{$F_{850}$\tablenotemark{d}}&
\colhead{$F_{450}$\tablenotemark{d}}&
\colhead{$A_V$\tablenotemark{d}}&
\colhead{Host\tablenotemark{e}}&
\colhead{Nearest\tablenotemark{e}}&
\colhead{Distance}\\
\colhead{}&
\colhead{(J2000.0)} &
\colhead{(J2000.0)} &
\colhead{}&
\colhead{(Jy/Bm)} &
\colhead{(Jy/Bm)} &
\colhead{mag}&
\colhead{Clump}&
\colhead{Clump}&
\colhead{(arcsec)}
}
\startdata
1727 & 21:44:43.0 & +47:46:43 & 0+I & 0.323 & 0.99 & 3.72 & 4 & 4 & 2.8\\
1729 & 21:44:48.3 & +47:44:59 & 0+I & 0.059 & 0.39 & 5.41 & 6 & 6 & 2.1\\
1731 & 21:44:51.7 & +47:40:44 & 0+I & 0.119 & 0.34 & 7.28 & 8 & 8 & 13.1\\
1732 & 21:44:52.0 & +47:40:30 & 0+I & 0.386 & 1.41 & 7.43 & 8 & 8 & 2.4\\
1736 & 21:44:53.9 & +47:45:43 & 0+I & -0.002 & 0.20 & 4.00 & - & 6 & 73.7\\
1737 & 21:44:57.0 & +47:41:52 & FLAT & 0.047 & 0.00 & 5.41 & 12 & 12 & 12.1\\
1739 & 21:45:02.6 & +47:33:07 & 0+I & 0.102 & 0.00 & 1.80 & 17 & 17 & 6.0\\
1741 & 21:45:08.3 & +47:33:05 & FLAT & 0.435 & 1.83 & 4.18 & 21 & 21 & 1.5\\
1742 & 21:45:27.8 & +47:45:50 & 0+I & -0.019 & -0.14 & 1.74 & - & 18 & 243.4\\
1743 & 21:45:31.2 & +47:36:21 & 0+I & 0.081 & -0.07 & 1.63 & 26 & 26 & 5.5\\
1746 & 21:45:58.5 & +47:36:01 & 0+I & 0.126 & 0.31 & 14.23 & 30 & 30 & 2.6\\
1753 & 21:47:03.0 & +47:33:14 & 0+I & 0.034 & 0.16 & 8.70 & 42 & 42 & 35.8\\
1754 & 21:47:06.0 & +47:39:39 & 0+I & 0.074 & 0.34 & 4.22 & 41 & 41 & 4.1\\
1757 & 21:47:21.2 & +47:32:26 & FLAT & 0.299 & 0.84 & 13.65 & 47 & 47 & 21.7\\
1758 & 21:47:22.6 & +47:32:05 & 0+I & 0.530 & 1.86 & 13.65 & 47 & 47 & 7.4\\
1759 & 21:47:22.7 & +47:32:14 & 0+I & 0.662 & 2.74 & 13.65 & 47 & 47 & 2.4\\
1760 & 21:47:28.7 & +47:32:17 & FLAT & 0.124 & 0.46 & 11.47 & 47 & 52 & 22.1\\
1763 & 21:47:55.6 & +47:37:11 & 0+I & 0.046 & 0.10 & 4.45 & 56 & 56 & 30.3\\
1764 & 21:47:59.2 & +47:36:26 & 0+I & 0.246 & 1.22 & 3.54 & 57 & 57 & 1.3\\
1765 & 21:52:14.3 & +47:14:54 & 0+I & 0.004 & -0.14 & 3.31 & - & 71 & 240.3\\
1776 & 21:52:37.7 & +47:14:38 & 0+I & 0.161 & 0.55 & 6.19 & 72 & 72 & 6.9\\
1788 & 21:53:06.9 & +47:14:34 & 0+I & 0.094 & 0.38 & 5.23 & 74 & 74 & 10.7\\
1798 & 21:53:25.0 & +47:16:22 & FLAT & 0.004 & 0.10 & 5.22 & - & 80 & 68.2\\
1803 & 21:53:28.3 & +47:15:43 & 0+I & -0.008 & -0.04 & 5.52 & - & 85 & 90.4\\
1804 & 21:53:28.6 & +47:15:51 & 0+I & 0.003 & 0.18 & 5.52 & - & 85 & 91.3\\
1811 & 21:53:30.4 & +47:13:10 & FLAT & -0.002 & 0.18 & 5.02 & - & 82 & 71.0\\
1814 & 21:53:31.4 & +47:22:17 & 0+I & 0.088 & 0.34 & 4.26 & 81 & 81 & 13.8\\
1818 & 21:53:33.0 & +47:14:39 & FLAT & 0.015 & 0.13 & 5.95 & 82 & 82 & 23.9\\
1820 & 21:53:33.1 & +47:14:18 & 0+I & 0.212 & 0.46 & 6.00 & 82 & 82 & 2.9\\
1823 & 21:53:33.9 & +47:17:24 & 0+I & 0.063 & 0.12 & 5.79 & 84 & 84 & 16.8\\
1826 & 21:53:34.7 & +47:20:44 & FLAT & 0.173 & 0.68 & 5.06 & 83 & 83 & 13.2\\
1830 & 21:53:36.3 & +47:19:03 & 0+I & 0.243 & 0.62 & 6.08 & 86 & 86 & 4.1\\
1831 & 21:53:37.0 & +47:18:17 & 0+I & 0.039 & -0.10 & 5.77 & 86 & 88 & 40.9\\
1833 & 21:53:38.3 & +47:19:35 & FLAT & 0.229 & 0.67 & 5.94 & 89 & 89 & 15.1\\
1841 & 21:53:42.0 & +47:14:26 & FLAT & 0.010 & -0.10 & 6.72 & - & 91 & 43.9\\
1846 & 21:53:49.7 & +47:13:51 & 0+I & 0.135 & 0.24 & 7.20 & 94 & 94 & 9.1\\
1849 & 21:53:55.7 & +47:20:30 & 0+I & 0.019 & -0.22 & 2.50 & - & 89 & 188.8\\
1851 & 21:53:58.1 & +47:14:45 & 0+I & 0.014 & 0.11 & 4.63 & 96 & 96 & 55.8\\
\enddata
\tablenotetext{a}{YSO observation designation from Dunham et al.\ (2105).}
\tablenotetext{b}{Location of YSO.}
\tablenotetext{c}{YSO Class (see text).}
\tablenotetext{d}{See text for definitions.}
\tablenotetext{e}{Clump observation designation (see Table 3).}
\end{deluxetable}

\clearpage
\begin{deluxetable}{ccccrrrrrr}
\tablecolumns{10}
\tablewidth{0pc}
\tabletypesize{\scriptsize}
\tablecaption{{\it Spitzer} Evolved Sources: Flux and Extinction\label{tab:disks}}
\tablehead{
\colhead{YSO Index\tablenotemark{a}}&
\colhead{R.A.\tablenotemark{b}}&
\colhead{Dec.\tablenotemark{b}}&
\colhead{Class Type\tablenotemark{c}}&
\colhead{$F_{850}$\tablenotemark{d}}&
\colhead{$F_{450}$\tablenotemark{d}}&
\colhead{$A_V$\tablenotemark{d}}&
\colhead{Host\tablenotemark{e}}&
\colhead{Nearest\tablenotemark{e}}&
\colhead{Distance}\\
\colhead{}&
\colhead{(J2000.0)} &
\colhead{(J2000.0)} &
\colhead{}&
\colhead{(Jy/Bm)} &
\colhead{(Jy/Bm)} &
\colhead{mag}&
\colhead{Clump}&
\colhead{Clump}&
\colhead{(arcsec)}
}
\startdata
1728 & 21:44:44.4 & +47:46:49 & II & 0.070 & 0.08 & 3.72 & 4 & 4 & 18.0\\
1730 & 21:44:49.1 & +47:46:21 & II & 0.003 & -0.09 & 4.07 & - & 4 & 66.7\\
1733 & 21:44:52.3 & +47:44:55 & II & -0.009 & 0.06 & 3.85 & - & 6 & 42.0\\
1734 & 21:44:53.2 & +47:46:07 & II & 0.004 & -0.05 & 3.37 & - & 6 & 86.1\\
1735 & 21:44:53.4 & +47:40:48 & II & 0.064 & 0.20 & 7.50 & 8 & 8 & 20.5\\
1738 & 21:45:01.5 & +47:41:22 & III & 0.007 & 0.22 & 6.23 & - & 12 & 46.8\\
1740 & 21:45:02.9 & +47:42:29 & III & 0.009 & -0.21 & 4.49 & - & 12 & 58.9\\
1744 & 21:45:41.0 & +47:37:07 & II & -0.014 & -0.00 & 3.81 & - & 28 & 103.3\\
1745 & 21:45:57.1 & +47:19:31 & II & -0.020 & 0.01 & 2.35 & - & 32 & 880.0\\
1747 & 21:46:05.7 & +47:20:48 & II & -0.004 & -0.14 & 1.01 & - & 32 & 787.2\\
1748 & 21:46:22.0 & +47:35:00 & III & -0.009 & -0.29 & 8.04 & - & 33 & 55.5\\
1749 & 21:46:26.6 & +47:44:15 & II & -0.006 & 0.05 & 2.15 & - & 37 & 361.9\\
1750 & 21:46:41.3 & +47:34:31 & II & -0.007 & -0.06 & 5.86 & - & 36 & 87.7\\
1751 & 21:46:46.7 & +47:35:55 & III & -0.019 & -0.08 & 3.31 & - & 36 & 92.2\\
1752 & 21:46:49.2 & +47:45:08 & III & -0.003 & 0.33 & 1.49 & - & 37 & 315.6\\
1755 & 21:47:10.8 & +47:32:04 & II & 0.022 & -0.05 & 14.19 & 43 & 43 & 20.0\\
1756 & 21:47:20.6 & +47:32:03 & II & 0.381 & 1.33 & 13.65 & 47 & 47 & 24.4\\
1761 & 21:47:32.9 & +47:32:08 & II & 0.041 & 0.12 & 10.98 & 52 & 52 & 27.5\\
1762 & 21:47:34.2 & +47:32:05 & II & 0.016 & -0.22 & 10.98 & 52 & 52 & 40.0\\
1766 & 21:52:19.2 & +47:15:55 & II & 0.004 & -0.02 & 4.15 & - & 72 & 209.8\\
1767 & 21:52:19.6 & +47:14:38 & II & -0.007 & 0.11 & 3.05 & - & 71 & 184.8\\
1768 & 21:52:30.7 & +47:14:06 & II & -0.000 & -0.01 & 4.22 & - & 71 & 73.2\\
1769 & 21:52:32.6 & +47:13:46 & II & -0.005 & 0.17 & 4.15 & - & 71 & 45.4\\
1770 & 21:52:32.7 & +47:14:09 & II & -0.011 & -0.03 & 4.15 & - & 71 & 63.2\\
1771 & 21:52:33.2 & +47:10:50 & II & 0.008 & -0.12 & 1.64 & - & 71 & 146.3\\
1772 & 21:52:34.0 & +47:13:43 & II & 0.042 & -0.56 & 4.02 & - & 71 & 34.1\\
1773 & 21:52:34.5 & +47:14:40 & II & 0.005 & 0.31 & 5.31 & - & 72 & 39.5\\
1774 & 21:52:36.1 & +47:11:56 & II & 0.003 & -0.23 & 2.62 & - & 71 & 78.1\\
1775 & 21:52:36.5 & +47:14:36 & II & 0.001 & -0.11 & 6.19 & 72 & 72 & 19.2\\
1777 & 21:52:39.7 & +47:11:10 & III & -0.017 & -0.04 & 2.77 & - & 71 & 130.3\\
1778 & 21:52:41.2 & +47:12:52 & II & 0.021 & -0.17 & 4.21 & - & 71 & 59.7\\
1779 & 21:52:42.7 & +47:12:09 & II & -0.001 & 0.17 & 3.86 & - & 71 & 96.1\\
1780 & 21:52:42.7 & +47:10:13 & II & -0.019 & 0.12 & 2.14 & - & 71 & 194.3\\
1781 & 21:52:45.4 & +47:10:39 & III & -0.002 & 0.14 & 2.55 & - & 71 & 183.5\\
1782 & 21:52:46.5 & +47:12:49 & II & 0.009 & 0.02 & 4.37 & - & 71 & 112.3\\
1783 & 21:52:49.5 & +47:12:17 & II & 0.026 & 0.00 & 4.02 & - & 71 & 151.2\\
1784 & 21:52:50.1 & +47:12:20 & II & -0.003 & 0.15 & 3.93 & - & 71 & 155.8\\
1785 & 21:52:57.0 & +47:15:22 & II & -0.009 & -0.07 & 4.48 & - & 73 & 101.3\\
1786 & 21:52:59.1 & +47:16:28 & III & -0.001 & -0.19 & 3.84 & - & 73 & 104.3\\
1787 & 21:53:03.9 & +47:23:30 & II & -0.005 & 0.24 & 2.79 & - & 79 & 270.8\\
1789 & 21:53:07.5 & +47:24:44 & II & 0.023 & -0.21 & 2.65 & - & 81 & 291.0\\
1790 & 21:53:08.4 & +47:12:53 & II & -0.008 & 0.02 & 3.77 & - & 76 & 89.5\\
1791 & 21:53:08.6 & +47:12:47 & II & -0.011 & 0.19 & 3.70 & - & 76 & 83.3\\
1792 & 21:53:18.4 & +47:14:20 & II & 0.038 & 0.15 & 4.81 & 75 & 75 & 70.1\\
1793 & 21:53:20.3 & +47:12:57 & II & -0.000 & -0.19 & 3.58 & - & 76 & 110.9\\
1794 & 21:53:22.6 & +47:28:17 & II & 0.005 & 0.09 & 1.06 & - & 81 & 384.3\\
1795 & 21:53:22.7 & +47:14:23 & II & 0.027 & -0.06 & 4.31 & - & 82 & 107.0\\
1796 & 21:53:23.0 & +47:14:08 & II & 0.013 & -0.08 & 4.13 & - & 82 & 103.9\\
1797 & 21:53:24.9 & +47:15:29 & II & 0.013 & -0.05 & 4.95 & - & 82 & 112.0\\
1799 & 21:53:26.3 & +47:15:43 & II & 0.004 & 0.01 & 4.95 & - & 80 & 108.5\\
1800 & 21:53:27.1 & +47:16:58 & III & -0.022 & -0.08 & 4.89 & - & 80 & 40.9\\
1801 & 21:53:27.3 & +47:14:50 & II & 0.005 & 0.23 & 5.01 & - & 82 & 69.2\\
1802 & 21:53:27.7 & +47:15:11 & III & -0.004 & 0.05 & 5.01 & - & 82 & 78.9\\
1805 & 21:53:28.8 & +47:16:13 & II & 0.019 & 0.30 & 5.97 & - & 80 & 88.1\\
1806 & 21:53:28.9 & +47:15:37 & II & -0.009 & -0.29 & 5.63 & - & 85 & 82.5\\
1807 & 21:53:28.9 & +47:16:52 & II & 0.018 & 0.23 & 5.69 & - & 80 & 57.9\\
1808 & 21:53:29.6 & +47:13:54 & II & -0.006 & -0.15 & 4.84 & - & 82 & 42.1\\
1809 & 21:53:30.2 & +47:16:03 & II & -0.020 & -0.03 & 5.97 & - & 85 & 85.0\\
1810 & 21:53:30.3 & +47:13:13 & II & -0.013 & -0.10 & 5.02 & - & 82 & 68.7\\
1812 & 21:53:30.5 & +47:14:44 & II & 0.001 & -0.04 & 5.12 & - & 82 & 39.7\\
1813 & 21:53:30.8 & +47:16:06 & II & -0.003 & 0.10 & 5.97 & - & 85 & 82.7\\
1815 & 21:53:31.6 & +47:16:28 & II & 0.008 & 0.14 & 6.11 & - & 84 & 71.4\\
1816 & 21:53:31.8 & +47:16:14 & II & -0.010 & -0.10 & 6.11 & - & 84 & 82.4\\
1817 & 21:53:32.0 & +47:16:03 & II & -0.009 & 0.01 & 6.07 & - & 85 & 72.4\\
1819 & 21:53:33.0 & +47:16:09 & III & 0.001 & 0.29 & 6.11 & - & 85 & 71.8\\
1821 & 21:53:33.2 & +47:13:41 & II & 0.074 & 0.15 & 6.00 & 82 & 82 & 34.2\\
1822 & 21:53:33.4 & +47:11:16 & II & -0.013 & 0.01 & 5.85 & - & 82 & 179.2\\
1824 & 21:53:34.0 & +47:15:55 & II & -0.014 & -0.16 & 6.11 & - & 85 & 54.6\\
1825 & 21:53:34.1 & +47:16:04 & III & -0.022 & -0.21 & 6.11 & - & 85 & 62.4\\
1827 & 21:53:35.7 & +47:12:26 & III & 0.020 & 0.12 & 7.00 & - & 87 & 109.7\\
1828 & 21:53:35.8 & +47:12:12 & II & 0.009 & 0.03 & 7.00 & - & 87 & 123.5\\
1829 & 21:53:36.2 & +47:10:27 & II & -0.012 & 0.09 & 5.79 & - & 93 & 193.2\\
1832 & 21:53:38.2 & +47:14:59 & II & 0.037 & 0.06 & 6.22 & 85 & 85 & 19.6\\
1834 & 21:53:38.4 & +47:12:31 & II & 0.021 & -0.11 & 7.47 & - & 87 & 105.6\\
1835 & 21:53:38.7 & +47:12:05 & II & 0.004 & 0.02 & 7.32 & - & 93 & 124.0\\
1836 & 21:53:38.9 & +47:13:27 & II & -0.004 & 0.17 & 7.49 & - & 87 & 53.1\\
1837 & 21:53:40.0 & +47:15:26 & II & 0.039 & -0.06 & 5.61 & 90 & 90 & 8.6\\
1838 & 21:53:40.4 & +47:15:08 & II & 0.031 & 0.08 & 6.22 & 90 & 90 & 19.7\\
1839 & 21:53:40.6 & +47:16:49 & II & -0.003 & 0.11 & 5.50 & - & 88 & 58.0\\
1840 & 21:53:40.8 & +47:15:42 & II & 0.003 & 0.19 & 5.61 & - & 90 & 14.8\\
1842 & 21:53:42.1 & +47:15:53 & II & -0.009 & -0.15 & 5.08 & - & 90 & 28.8\\
1843 & 21:53:42.5 & +47:18:25 & II & 0.039 & 0.05 & 5.56 & - & 92 & 59.5\\
1844 & 21:53:45.2 & +47:12:35 & II & -0.000 & 0.24 & 5.81 & 93 & 93 & 54.5\\
1845 & 21:53:46.5 & +47:14:35 & II & 0.021 & 0.05 & 6.38 & - & 91 & 39.9\\
1847 & 21:53:51.6 & +47:16:48 & II & -0.002 & -0.06 & 4.06 & - & 92 & 75.9\\
1848 & 21:53:51.8 & +47:07:11 & III & -0.017 & 0.06 & 0.85 & - & 93 & 322.4\\
1850 & 21:53:57.9 & +47:19:34 & II & -0.018 & 0.02 & 2.82 & - & 92 & 170.0\\
1852 & 21:54:00.3 & +47:25:22 & II & 0.020 & -0.14 & 1.09 & - & 81 & 354.7\\
1853 & 21:54:01.4 & +47:14:21 & II & 0.016 & -0.10 & 4.91 & 96 & 96 & 62.5\\
1854 & 21:54:06.0 & +47:22:05 & II & 0.005 & -0.19 & 1.37 & - & 89 & 319.6\\
1855 & 21:54:08.7 & +47:13:57 & II & 0.000 & 0.25 & 3.39 & - & 96 & 130.9\\
1856 & 21:54:12.5 & +47:14:35 & II & -0.023 & -0.01 & 1.84 & - & 96 & 174.4\\
1857 & 21:54:18.7 & +47:12:09 & II & 0.007 & -0.07 & 1.43 & - & 95 & 255.2\\
\enddata
\tablenotetext{a}{YSO observation designation from Dunham et al.\ (2105).}
\tablenotetext{b}{Location of YSO.}
\tablenotetext{c}{YSO Class (see text).}
\tablenotetext{d}{See text for definitions.}
\tablenotetext{e}{Clump observation designation (see Table 3).}
\end{deluxetable}

Figure \ref{fig:cummass} compares 
the cumulative mass, as determined from the extinction map within the {\it Spitzer} footprint, 
to the emergence of these YSOs as a function of the underlying extinction. It is clear that the YSOs are
at least as biased to higher extinction regions as the clump mass in the cloud, reinforcing the notion that
stars form where there is significant material \citep[e.g.,][]{jdk04,Hatchell05,kea06,Konyves13}.
Furthermore, a large fraction of the
most embedded YSOs, the protostellar Class 0/I and Flat, are found at extreme extinctions of $A_V > 5$, 
which is also where the SCUBA-2-derived mass resides. The more-evolved Class II+III 
sources  are the least biased toward high extinction. This progression of Classes against extinction
in IC\,5146 can be interpreted as an evolutionary sequence. The heavily embedded Class 0/I and 
Flat sources still reside near their natal material while the older Class II and Class III sources
have had an opportunity to disperse or move away from their remnant birth sites. 

\begin{figure}[htb]
\includegraphics[scale=0.4]{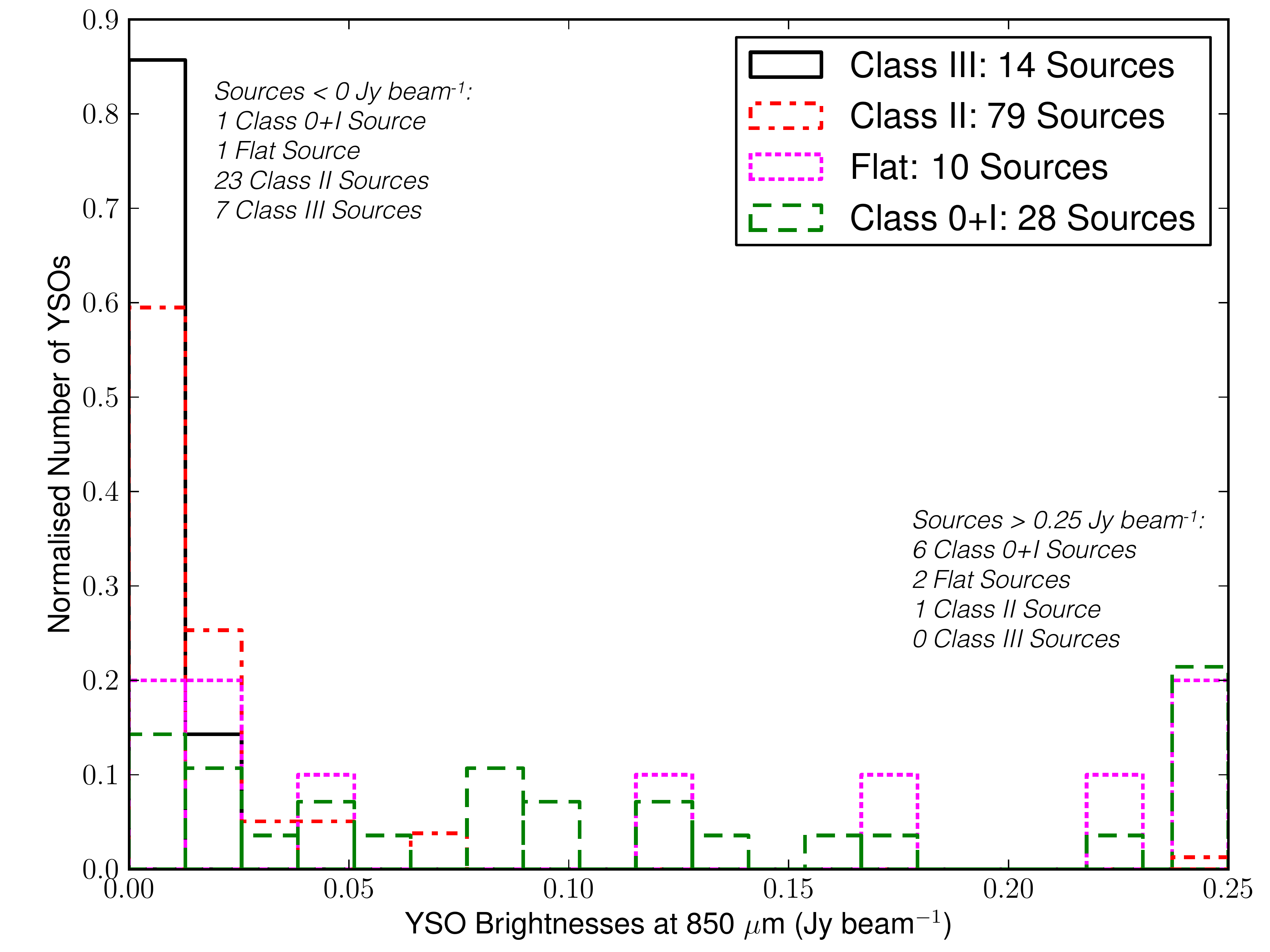}
\caption{
Histogram of the 850\,$\mu$m SCUBA-2 flux at the position of {\it Spitzer}
YSOs normalized for each class (highlighted as in the legend).
All 131 sources can be found in Table \ref{tab:clumps_obs}.
}
\label{fig:yso_850}
\end{figure}

Figure \ref{fig:yso_850} shows and Tables \ref{tab:protos} and \ref{tab:disks} record the distribution of 
submillimetre flux densities associated with the position of each YSO 
from the 850\,$\mu$m maps. Most of the Class II and Class III YSOs are associated with little to no submillimetre 
emission  at either 450\,$\mu$m or 850\,$\mu$m. This result is as expected \citep{dea15} since those YSOs
are more evolved, with small or non-existent surrounding envelopes and enough time has passed for 
them to have moved from their natal surroundings. The younger Class 0/I and Flat YSOs are seen to be
embedded in regions of higher flux emission, appropriate for dense envelopes and/or
proximity to material responsible for their birth. While there appear to be a subset of Class 0/I and Flat
YSOs without significant coincident submillimetre emission, it is likely that these sources have been 
mis-classified and are actually somewhat older. For example, \citet{he15} surveyed 546 such {\it Spitzer}-classified YSOs
using HCO$^+(J=3-2)$ and found that only 84\% were bona fide young protostars. Additionally, using
SCUBA-2 observations in Southern Orion A, \citet{mea16} find a similar fraction of likely mis-identified
Class 0/I and Flat sources.

\begin{figure}[htb]
\includegraphics[scale=0.5]{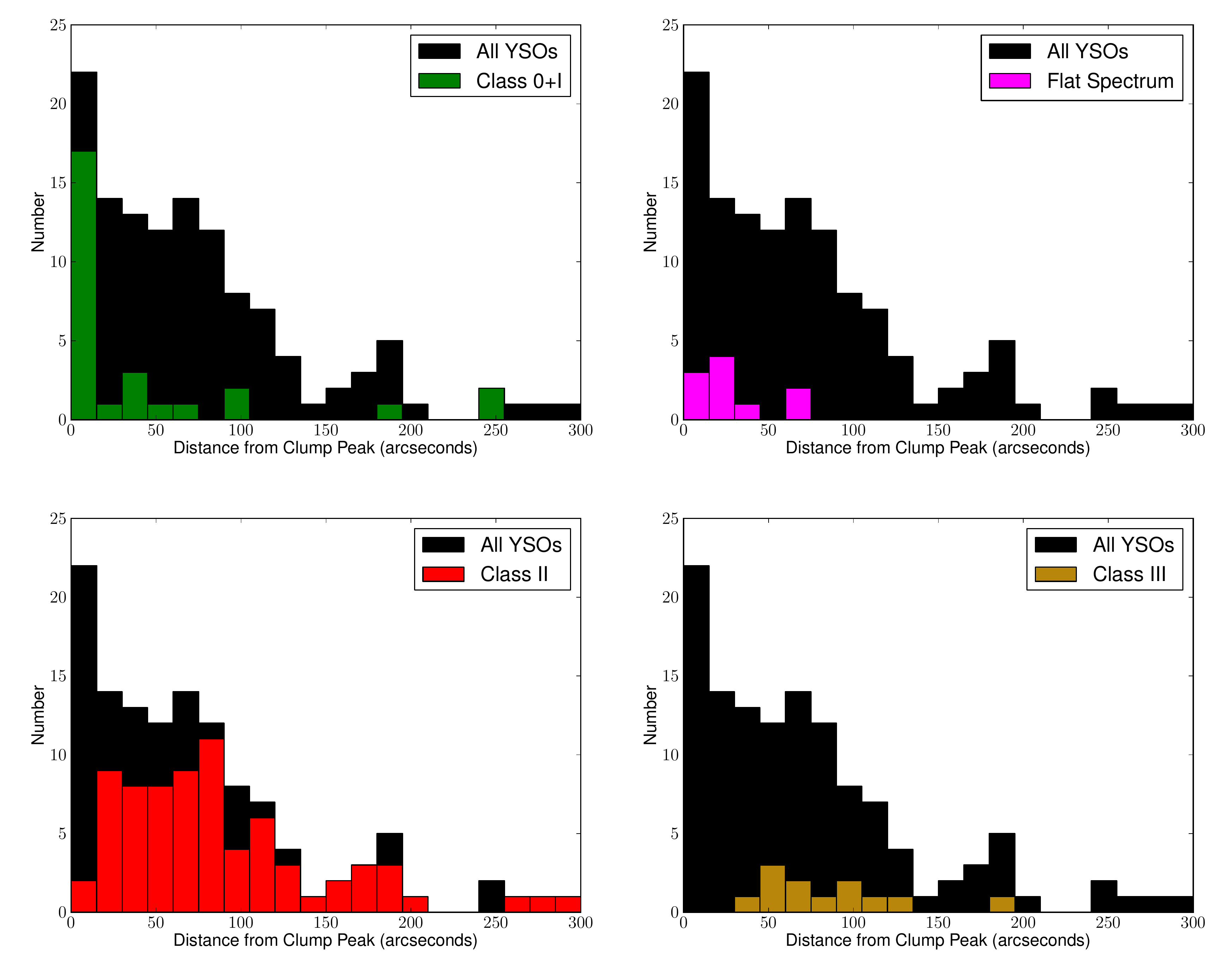}
\caption{
Histograms of YSO angular offset from nearest clump peak. In all panels, black represents the entire sample of YSOs within
300\arcsec of a clump peak.
(a) Offsets for Class 0/I sources (Green). (b) Offsets for FLAT sources (Pink). (c) Offsets for Class II sources (Red). (d) 
Offsets for Class III sources (Gold). See the text for details on YSO class determination. 
}
\label{fig:yso_dist}
\end{figure}

Figure\ \ref{fig:yso_dist} shows the distribution of YSO angular offsets from clump peaks. The
most deeply embedded Class 0/I YSOs are found to dominate the distribution at very small offsets,
$d < 15\arcsec$ ($1.4 \times 10^3\,$AU at a distance of 950\,pc) roughly equivalent to the 850\,$\mu$m beam.
The roughly 25\% of the Class 0/I's that are found at larger offsets from the clump peaks are likely mis-identified
as discussed in the previous paragraph. Furthermore, the ten Flat spectrum YSOs are all found within $80\arcsec$ of 
a clump peak while the Class II YSOs are predominantly found at intermediate offsets, 
$d < 150\arcsec$ ($1.4 \times 10^4\,$AU), 
but with a very broad overall range.  The few Class III sources are almost exclusively found at a 
 similar intermediate offsets, $40\arcsec < d < 180\arcsec$, although three lie further than $300\arcsec$ away
 from any clump peak.

\begin{figure}[htb]
\includegraphics[scale=0.55]{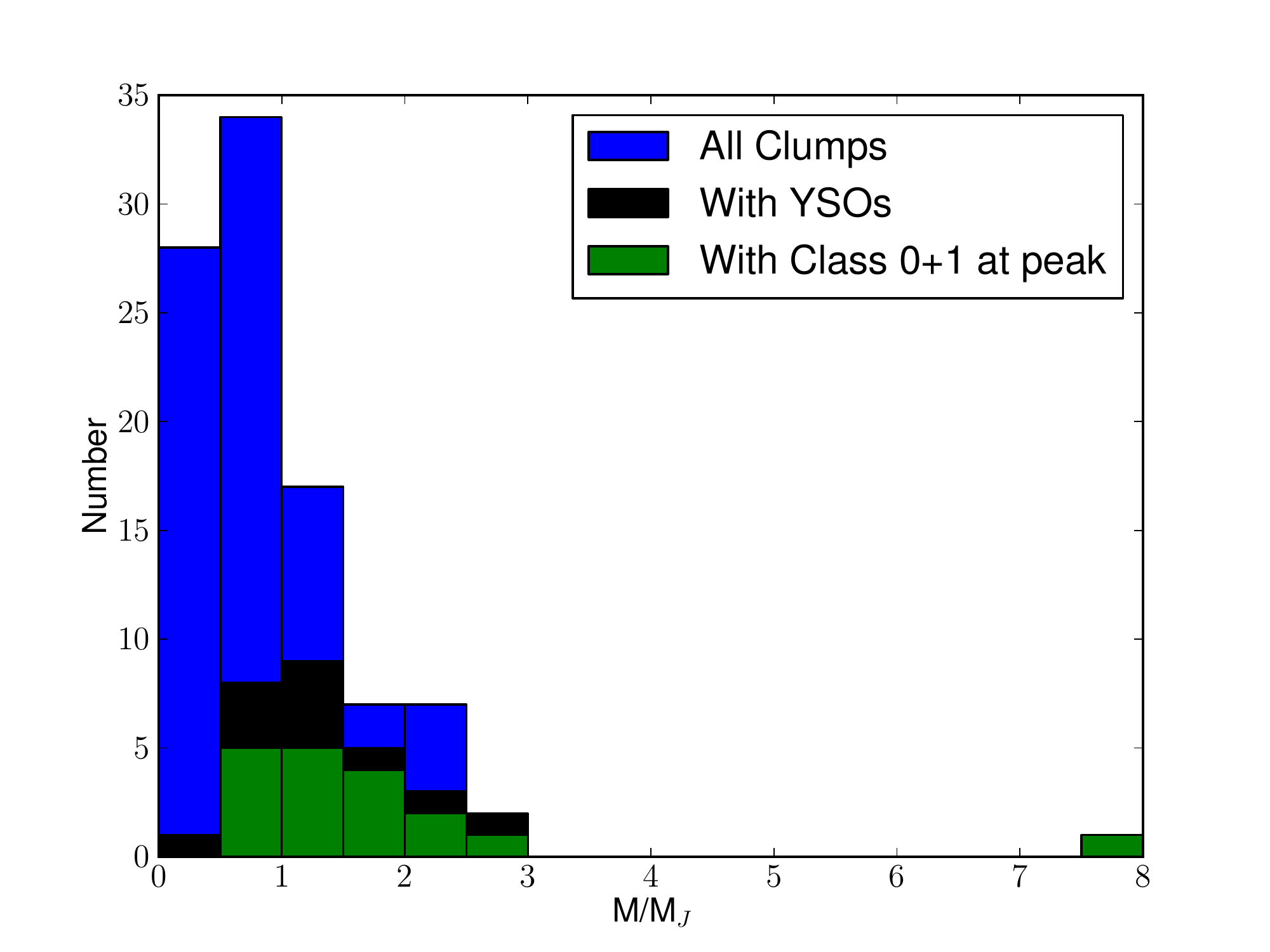}
\caption{
Histogram showing the Jeans stability, $M/M_J$, for all 96 clumps, the subset of 29 clumps that harbour YSOs, and the subset with protostars near their peak. The fraction of clumps harbouring YSO increases with $M/M_J$. Of the clumps harbouring YSOs, the fraction with protostars located near the clump centre increases as well with $M/M_J$.
}
\label{fig:yso_mj_hist}
\end{figure}

\begin{figure}[htb]
\includegraphics[scale=0.55]{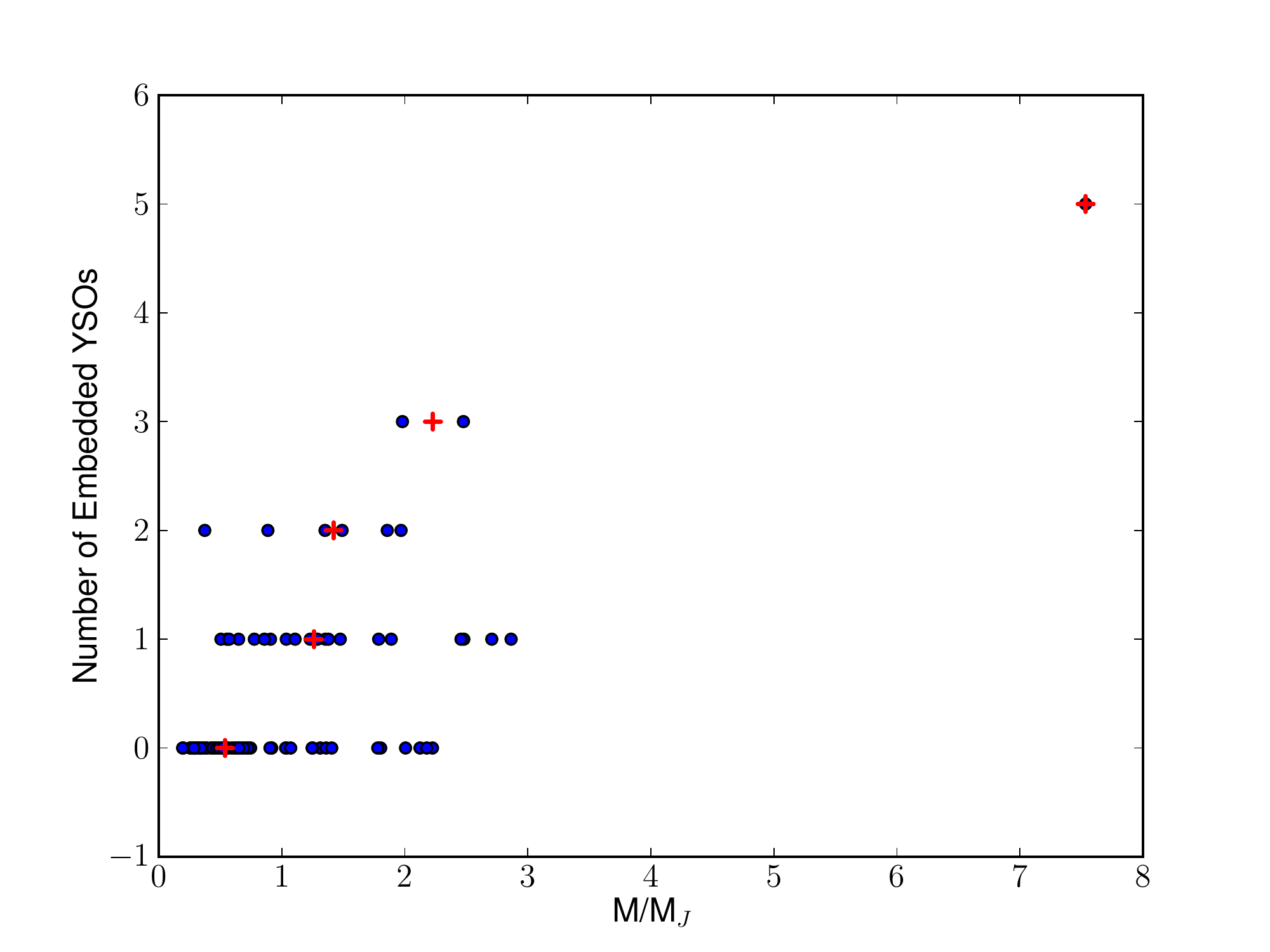}
\caption{
Scatter diagram for clumps showing the Jeans stability measure, $M/M_J$, versus the number of embedded YSOs. Denoted by a red star are the median $M/M_J$ as a function of embedded YSOs. There is a clear relation revealed, with clumps that harbour a greater number of YSOs also having a larger median $M/M_J$.
}
\label{fig:yso_mj}
\end{figure}

\begin{figure}[htb]
\includegraphics[scale=0.55]{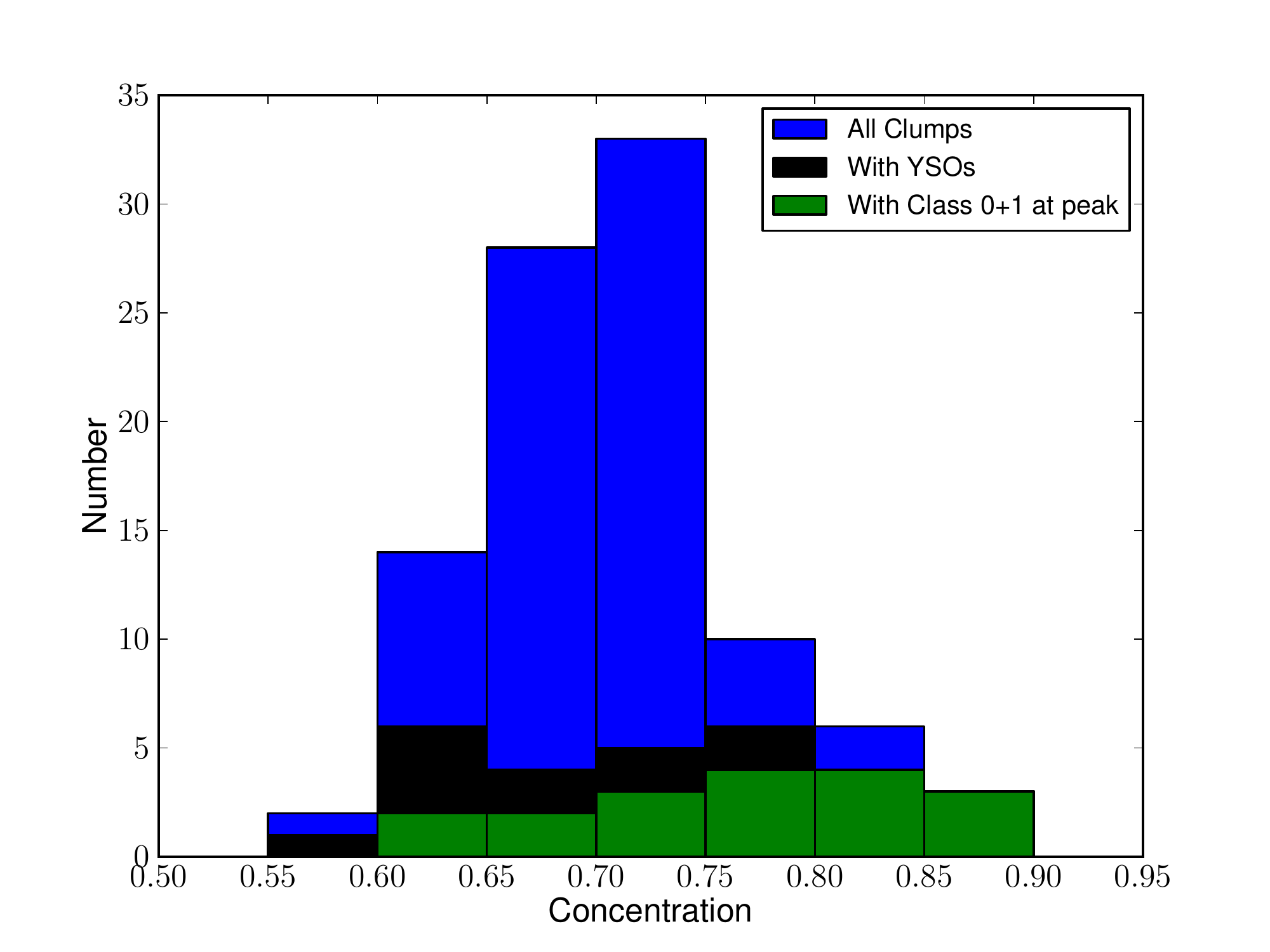}
\caption{
Histogram showing the concentration measure (Eqn.\ \ref{eqn:c}) for all 96 clumps, as well as the subset of 29 clumps that harbour YSOs and those 18 with protostars near their peak. Although clumps embedded YSOs span the entire range of concentration, the clumps with higher concentration are most likely to have protostars near their peak.}
\label{fig:yso_conc_hist}
\end{figure}

Detailed consideration of the physical properties of the clumps containing YSOs reveals a few 
important trends. Figure\ \ref{fig:yso_mj_hist} shows histograms of the Jeans stability ratio for all 96 clumps, and just those with YSOs.
Clumps harbouring YSOs within their boundary
have a distribution of Jeans stability ratio that is skewed toward larger (more unstable) values 
and those harbouring a protostar near the peak makes this skew slightly more extreme.
Figure\ \ref{fig:yso_mj} shows the Jeans stability versus the number of embedded YSOs.
The degree of instability is in general larger for clumps that contain a greater number of embedded YSOs.
Finally, Figure\ \ref{fig:yso_conc_hist} shows histograms of concentration for all clumps and those with YSOs.
Clumps across the entire range of observed concentrations contain YSOs, with
YSO-harbouring clumps dominating both the low and 
high concentration ends of the distribution. Considering only those clumps with protostars close to
their peaks, however, one finds an enhancement toward high concentration.

\subsection{Clustering of Sources}
\label{sec:anal:cluster}

Given that there are clear differences between the spatial distribution of the clumps and protostars
in the Cocoon Nebula and the Northern Streamer, we have analyzed the clustering properties of these 
objects.  We calculate the local surface density of clumps around each clump
by using the separation to the Nth nearest neighbour, i.e.,
\begin{equation}
\sigma_{N} = \frac{N}{\pi r_N^2}
\end{equation}
where $\sigma_N$ is the local surface density, and $r_N$ is the distance to the Nth nearest
neighbour.  The fractional uncertainty of the surface density is $N^{-0.5}$
\citep{Casertano85, Gutermuth09}.  Following \citet{Kirk16b} and \citet{Lane16}, we
calculate the surface density values for 5 and 10 nearest neighbours, finding no significant difference
between the two approaches.  We use a similar
approach to calculate the local surface density for each protostar.  For the protostar
calculation, we consider the Class 0/I and Flat sources as a single group and the Class II
sources as a second group, using only the distances to other sources in the same group
for the surface density calculation, i.e., the Class II measurement only considers the 
distances between Class II sources.

\begin{figure}[htb]
\includegraphics[scale=0.48]{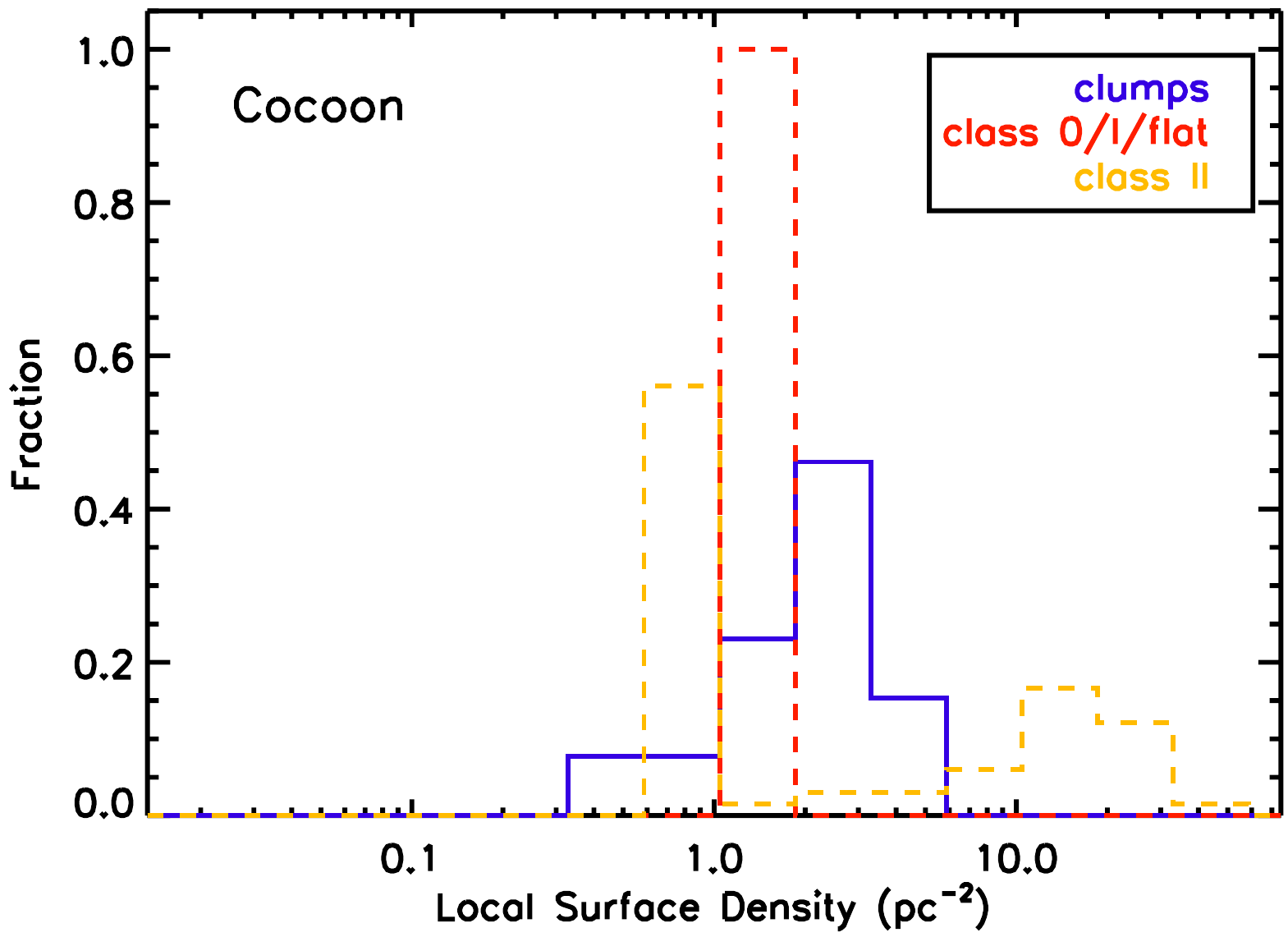}
\medskip
\includegraphics[scale=0.48]{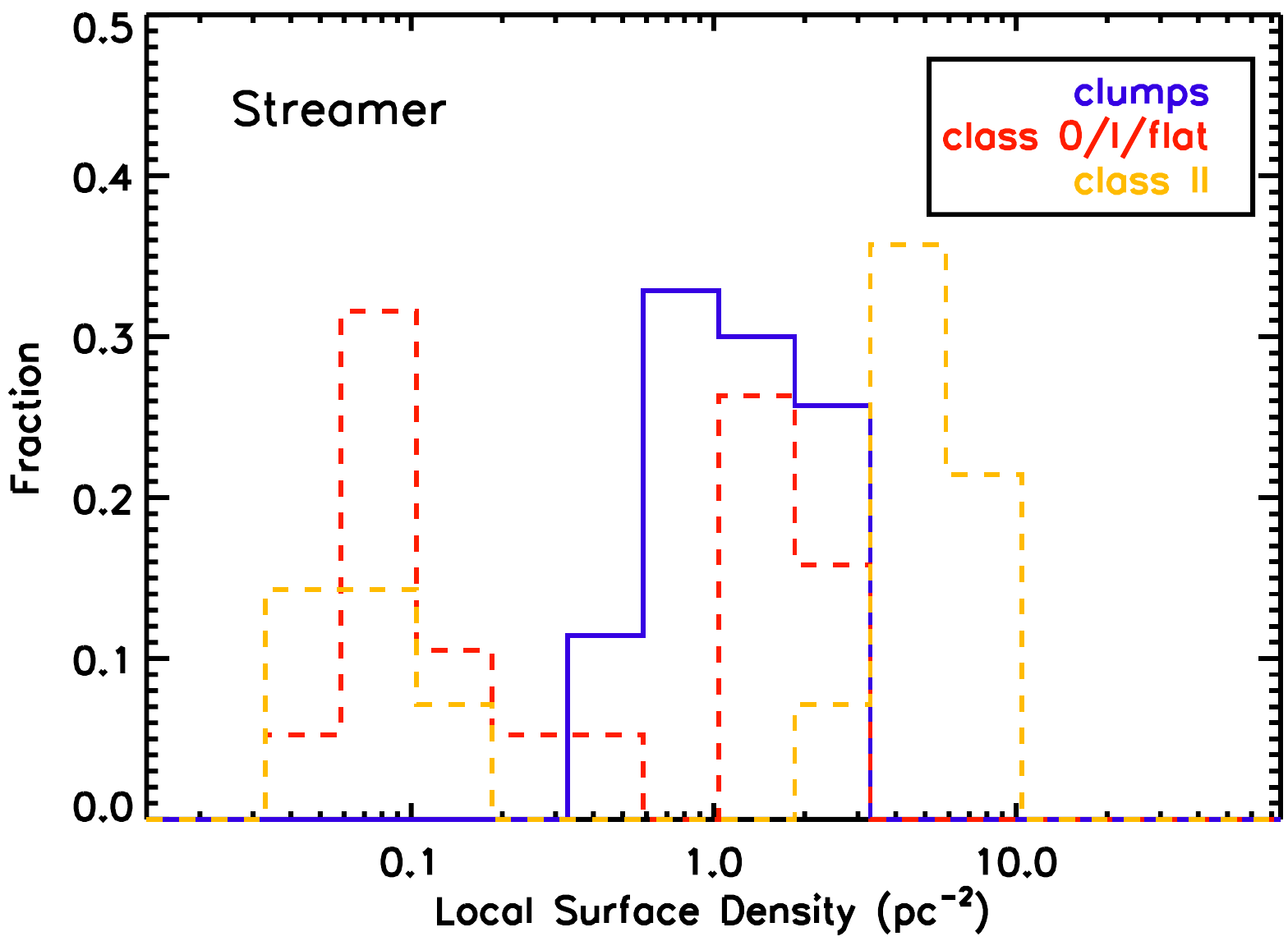}
\caption{The distribution of local source surface densities derived for the Cocoon
	Nebula (left) and the Northern Streamer (right).  In each panel, the solid blue line
	shows the clumps, while the dashed red line shows the Class 0/I and flat
	protostars and the dashed yellow line shows the Class II protostars.
	Results are plotted here for surface densities calculated using the ten
	nearest neighbours; using instead the five nearest neighbours gives
	a similar result.}
\label{fig_surfdens}
\end{figure}

Figure~\ref{fig_surfdens} shows the distributions of the surface densities derived for
sources in the Cocoon Nebula and Northern Streamer separately.  The most obvious
difference between the two subregions is in the protostellar surface densities, with
protostars in the Cocoon Nebula showing much stronger signs of clustering than the
Northern Streamer, confirming the visual impression of the two regions.  Clumps in
the Cocoon Nebula also appear to have a slight tendency to lie closer together than
in the Northern Streamer, although this is less visually apparent.  We ran 
two-sided Kolmogorov-Smirnov tests on the two distributions, and found probabilities
of 8\% or less that the Cocoon and Streamer populations are drawn from the same
parent sample (0.45\%, 0.06\%, and 8\% for the clumps, Class 0/I/Flat and Class II
sources respectively).  We also ran a Mann-Whitney test to examine the probability that
sources in the Cocoon tend to have higher surface densities than sources in the Streamer 
and find significant probabilities for the clumps and Class 0/I/Flat sources (probabilities are 
99.99\%, 98\%, and 87\% for the clumps, Class 0/I/Flat, and Class II sources, respectively).

\begin{figure}[htb]
\includegraphics[scale=0.7]{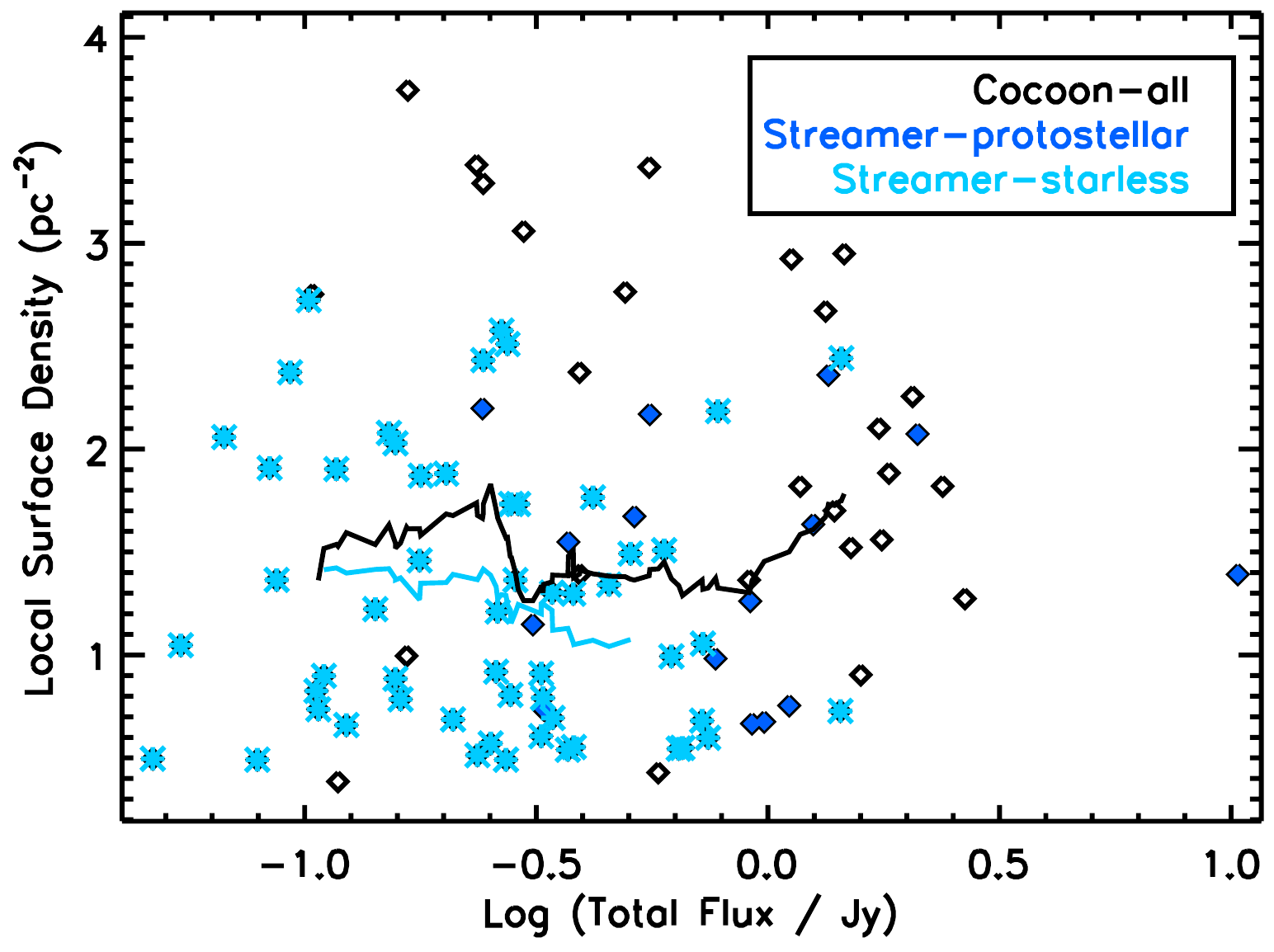}
\caption{A comparison of the total flux and local surface densities of the clumps
	in IC\,5146.    Northern Streamer
	protostellar clumps are shown in dark blue filled diamonds while the starless clumps are
	shown in light blue asterisks}.  The solid black and light blue lines show the co-moving
	window mean values of the surface density for the full sample (black) and
	the starless clumps in the Northern Streamer only (light blue).  No clear
	trend is seen between the surface density and clump total flux.
\label{fig_sd_totflux}
\end{figure}

Following \citet{Kirk16b} and \citet{Lane16}, we use a second clustering measurement
adapted from protostellar analysis in simulations by 
\citet{Maschberger11}, and we compare the local surface 
density of clumps to their total flux.  In the Orion A and B molecular clouds, 
\citet{Lane16} and \citet{Kirk16b} found strong evidence of a tendency for higher
flux sources to inhabit locally higher surface density environments, which was interpreted,
in combination with other lines of evidence, as a sign of dense core mass segregation
in those clouds.  We run a similar comparison here.  Figure~\ref{fig_sd_totflux} shows
the local surface density and flux for the clumps in IC\,5146.  We analyze both the 
full sample of clumps, and separately, only the clumps in the Northern Streamer.
We did not expect to see signs of mass segregation within the Cocoon Nebula, given
that the clumps are roughly arranged in a ring around the protostellar cluster, suggesting
that the region is more evolved, and that the submillimetre clumps may represent a second round
of star formation.  Even if mass segregation were present, the ring-like geometry of
the clumps would make it difficult to measure in the Cocoon Nebula.  
Even removing the Cocoon Nebula
sources, however, we see no signs of a trend of increasing local surface density
with higher flux clumps.  \citet{Kirk16b} and \citet{Lane16} ran their analysis
considering both the full sample of dense cores in Orion and also only the starless cores, 
since the presence of a protostar could imply a higher temperature, thus increasing 
the flux of a dense core
without also implying an increased mass.  No trend is seen in Figure~\ref{fig_sd_totflux}
for either the full set of clumps or only starless clumps within the Northern Streamer.
To highlight this lack of a trend, we also show the co-moving mean surface density
values for both the full set of IC\,5146 clumps and the starless clumps within the
Northern Streamer, using a smoothing window width of 20 clumps.

The lack of support for the presence of mass segregation within the Northern Streamer
is surprising, given the strong signs of core mass segregation found in the Orion
A and B molecular clouds using a similar analysis and identical observing technique with 
the JCMT.  There are two possible explanations for the different behaviour in IC\,5146.
First, mass segregation in dense cores and clumps may not be universal, but might
be restricted only to the most clustered systems like Orion.  This possibility can soon
be tested using the data available for other clouds observed by the JCMT GBS team.
A second possibility is that the difference is due to the much larger distance to IC\,5146 than
Orion (and other GBS clouds).  At the adopted 950~pc, IC\,5146 is too distant to resolve
individual dense cores, and we instead are only sensitive to the larger clump emission
structures.  The local surface density values derived for the clumps already suggests
that we may be missing important details in un-resolved substructures.  Within the
starless clumps in the Northern Streamer, the range of local surface densities is
$0.49$~pc$^{-2} \le \sigma_{10} \le 2.7$~pc$^{-2}$, a factor of only 5.5.  In 
Orion~A, \citet{Lane16} find a range in surface densities of more than a factor of 100,
and in Orion~B, the range is about a factor of 60.  
In IC~5146, the mean clump radius is 0.15~pc, whereas in Orion~A and B, the FellWalker-based
dense cores have mean effective radii of 0.06~pc.
The filamentary geometry of the Northern Streamer may also make it more difficult to 
measure clustering properties, however, the recent work by \citet{Nunes16} suggests
that with sufficient resolution, clusters of dense cores might be expected.  
Using WISE data, \citet{Nunes16}
identify five new candidate protostellar clusters in the area around the Northern Streamer.
Three of these, NBB~2, 3, and 4, all lie within areas we mapped with SCUBA-2, and two of
these, NBB~2 and 4, lie close to each of the two brightest submillimetre clumps within this
area.

It will be important to measure
the clustering properties in other nearby molecular clouds to test whether IC~5146
appears different merely because of its distance.

\section{Discussion}
\label{sec:disc}

The contrast between the Cocoon Nebula and the
Northern Streamer is evident throughout this analysis,
revealing the large variation in structure and star formation that can arise
within a single molecular cloud. Both regions show clear
fragmentation of bright submillimetre emission into dense clumps with a similar fraction
of the extinction-derived mass observable at 850\,$\mu$m (~5\% for the Cocoon Nebula 
versus 4\% for the Northern Streamer).
The Northern Streamer, seen in extinction as a network of parallel filaments, breaks
into chains of clumps when observed in the submillimetre. While only a small fraction of 
these clumps host YSOs, the majority of YSOs in 
the region lie within the clumps.  The Cocoon Nebula, in contrast, appears more as a 
broken, fragmented ring surrounding a cluster of predominantly Class II protostars. A
large fraction of these Cocoon Nebula clump fragments host YSOs but only a small
fraction of the YSO population remains embedded within clumps. This analysis is 
independent of YSO evolutionary stage.

The Cocoon Nebula is situated at the eastern end of the Northern Streamer and
thus we theorize that it is a more-evolved portion of IC\,5146,
having produced 65 Class II YSOs versus the 14 Class II YSOs found in the
Northern Streamer. The number of Class 0/I and Flat sources is identical
between the two regions despite the Northern Streamer containing twice the total
extinction-derived mass of the Cocoon Nebula and almost three times as many submillimetre
clumps. The variation in evolutionary age between the two regions may be due to the more 
localized mass concentration in the Cocoon Nebula, suggesting a higher initial density leading to a 
faster collapse.  Alternatively, it might simply be due to the known enhanced
gravitational instability produced at the ends of cylindrical structures \citep{pea11, pea12},
assuming that the entire IC\,5146 might have begun as a single, elongated entity.

Early in its existence, the Cocoon Nebula developed a blister cavity 
due to the massive star BD+46$^{\circ}$ 3474, evacuating the central region and spreading material 
outward. This energetic event appears to have compressed the remaining
clumps into a ring and triggered a second round of star formation, perhaps explaining the
high fraction of current clumps in the Cocoon with embedded YSOs. As suggested by the
collect and collapse model
\citep{Whitworth94a}, the compression may also 
explain the higher clump masses observed in the Cocoon Nebula 
although there is only a modest
difference in the mean densities within the Cocoon Nebula versus the Northern Streamer,
$410\,M_\odot\,{\rm pc}^{-3}$ versus $350\,M_\odot\,{\rm pc}^{-3}$ respectively.
Less clear is an explanation for
the almost constant fraction of extinction-derived mass observed at 850\,$\mu$m across
both the Cocoon Nebula and the Northern Streamer. If this fraction traces clump formation
efficiency then one might naively expect that the recent compression in the Cocoon Nebula should 
result in a higher fractional value than that found in the Northern Streamer. Alternatively, if
star formation is ramping down in the Cocoon Nebula then the anticipated fractional
value would be lower than in the Northern Streamer as more of the dense gas has been
locked into stars or disrupted. Perhaps the similarity in the dense gas fraction
is due to a combination of these two effects, with the enhancement and dispersal roughly
cancelling out.

The Northern Streamer, conversely, has relatively few of the Class II YSOs (18\%) despite
being surrounded by 67\% of the extinction-derived mass of the entire IC\,5146 cloud 
and accounting for 57\% of the submillimetre-derived mass. This suggests that
star formation has not yet ramped up to the level found in the Cocoon Nebula.
The lower mean (and median) masses for the clumps in the Northern Streamer may
also be evidence of youth. Nevertheless, both the Jeans mass and concentration measures for
the clumps in the Northern Streamer suggest that it is poised for significant star formation activity
continuing into the future. One additional indirect piece of evidence that star formation is only now
ramping up in the Northern Streamer is found in the nature of Clump 47 and its nearby environment.
Found near the eastern end of one of the Northern Streamer filaments, Clump 47 is the single most
massive clump observed in the IC\,5146 region. It is extremely Jeans unstable, assuming only thermal 
pressure support, and hosts a small ensemble of protostars already. The conditions within this clump
may be similar, though scaled down, from those assumed to have led to the blister HII region associated
with the Cocoon Nebula, including the notion that gravitational instabilities are associated with the
ends of filaments.

\section{Summary}
\label{sec:conc}

In this paper we have presented a first look at IC\,5146 from JCMT SCUBA-2 dust continuum data. 
We are able to confirm a number of key properties regarding IC\,5146 as listed below:
\begin{itemize}
\item The Cocoon Nebula is a site of spherical, clustered star formation
that was disrupted by the energetic outflows of the massive BD+46$^{\circ}$
3474. The 850\,$\mu$m emission observed in this region shows a shell-like
geometry around the central cavity and the clumps are often found to harbour protostars,
suggesting vigorous but non-sustainable star formation. 
The Northern Streamer in contrast is an elongated network of near-parallel
filamentary structure undergoing modest star formation activity. Both regions
reveal fragmentation of the denser material, leading to localized sites of
star formation.
\item Dense clumps are associated with regions of high 
extinction, with over 50\% of the submillimetre emission
observed by SCUBA-2 in regions of high extinction ($A_V > 5$). The
total estimate of the extinction-derived mass is $\sim 15600\,M_\odot$ whereas
the SCUBA-2 dust continuum mass is only $\sim630, M_{\odot}$. Both subregions within IC\,5146
have converted approximately the same fraction of extinction-derived mass into 
submillimetre-visible mass.
\item Using the FellWalker clump-finding algorithm, 96 clumps are identified
in a mass range of $0.5-120\,M_{\odot}$.  A majority of the clumps
are found to be Jeans stable and one-third are found to harbour YSOs. Those
clumps that harbour protostars are typically the most Jeans unstable. Furthermore,
the clumps in the Cocoon Nebula are both more massive and more Jeans unstable. 
\item 
In general, the YSOs and submillimetre dust emission are both found to be strongly associated 
with the higher extinction regions within the cloud. Younger YSOs, Class 0/I and Flat, are 
found closer to the centres of clumps and at higher submillimetre flux density levels than
the more-evolved Class II and Class III objects.
\end{itemize}

\acknowledgments{

We thank the anonymous referee for constructive comments that added to the clarity of the paper.
We also thank Laurent Cambr\'esy for providing the extinction data used in this analysis and
Michael Dunham for allowing us to utilize his accumulated
{\it Spitzer} catalog information on the IC\,5146 region (Dunham et al. 2015) before publication.
Doug Johnstone is supported by the National Research Council of Canada and by an NSERC Discovery Grant. 
Steve Mairs was partially supported by the Natural Sciences and
Engineering Research Council (NSERC) of Canada graduate scholarship
program.
SC, DJ, and HK thank Herzberg Astrophysics at the National Research Council of Canada
for making this project possible through their co-op program.

The authors wish to recognise and acknowledge the very significant cultural role and reverence that the summit of 
Maunakea has always had within the indigenous Hawaiian community. We are most fortunate to have the opportunity 
to conduct observations from this mountain. The James Clerk Maxwell Telescope has historically been operated by the 
Joint Astronomy Centre on behalf of the Science and Technology Facilities Council of the United Kingdom, the 
National Research Council of Canada and the Netherlands Organisation for Scientific Research. Additional funds for 
the construction of SCUBA-2 were provided by the Canada Foundation for Innovation. The identification number for
 the programme under which the SCUBA-2 data used in this paper is MJLSG36. The authors thank the JCMT 
 staff for their support of the GBS team in data collection and reduction efforts. 
The Starlink software \citep{Currie14} is supported by 
the East Asian Observatory.  These data were reduced using a development version from 
December 2014 (version 516b455a).
This research has made use of NASA's
 Astrophysics Data System and the facilities of the Canadian Astronomy Data Centre operated by the National 
 Research Council of Canada with the support of the Canadian Space Agency. This research used the services of 
 the Canadian Advanced Network for Astronomy Research (CANFAR) which in turn is supported by CANARIE, 
 Compute Canada, University of Victoria, the National Research Council of Canada, and the Canadian Space Agency.
 This publication makes use of data products from the Two Micron All Sky Survey, which is a joint project of the University of Massachusetts and the Infrared Processing and Analysis Center/California Institute of Technology, funded by the National Aeronautics and Space Administration and the National Science Foundation. 
This research makes use of {\sc{APLpy}}, an open-source plotting package for Python hosted at http://aplpy.github.com \citep{Robitaille12}, and {\sc{matplotlib}}, a 2D plotting library for Python \citep{matplotlib}.

\facility{JCMT (SCUBA-2)}
\software{IRAF, Starlink \citep{Currie14}, SMURF \citep{jenness2013,cea13a, cea13b}, CUPID \citep{bea07,berry2013}, APLpy, Matplotlib}

\appendix
\section{Investigation into the Dust Temperature in IC\,5146}

The temperature of the dust within IC\,5146 can be estimated by comparing the flux density emitted
at 450\,$\mu$m and 850\,$\mu$m.  A careful comparison requires not only consideration
of the differing beamsizes at each wavelength, but also proper treatment of the
full beam profile, which, for the JCMT, includes a significant amount of flux
in the `secondary' beam at 450~$\mu$m \citep{dea08, dea13}.  Using the procedure
outlined in \citet{pea15} and \citet{rea15}, Rumble et al.~(in prep)
compiled temperature estimates for all of the GBS clouds based on the Legacy Release 1
reductions \citep[see \S 3.1 and][]{mea15}.   Here, we examine the temperature map derived for IC\,5146 to 
determine whether or not our assumption of a constant temperature of 15~K for the clumps 
is reasonable.

The Rumble et al.~(in prep) temperature maps exclude pixels for which the value derived
for the temperature is less than or equal to the error estimated for that temperature.
Since much of the area observed in IC\,5146 has relatively low flux densities, particularly
at 450\,$\mu$m, a large fraction of the map does not have reliable temperature 
estimates.  We first examine all pixels in the temperature map where a temperature
is measured.  From this analysis, we find a mean temperature of 14$\pm$5~K.  
Temperatures are similar for both the Northern Streamer and Cocoon Nebula subregions, 
with mean temperatures of 14$\pm$5~K and 14$\pm$6~K in the former and latter 
regions, respectively. Of the pixels with measureable temperatures, roughly
40\% lie in the Northern Streamer and 60\% lie in the Cocoon Nebula.

We next examine the temperature estimates at the locations of the clumps 
(see \S \ref{sec:anal:clumps}).
Using either the average temperature across the clump, or only the temperature
at the location of the peak 850\,$\mu$m flux density, we find that the clump dust has a similar
temperature to the region-wide measures above.  The mean temperature of clumps
calculated using the full clump extent is 14$\pm$3~K, while the mean temperature
of clumps using the temperature only at the clump peak is 13$\pm$2~K.  There is 
no significant difference in these values if the clumps are considered separately
within the Northern Streamer and Cocoon Nebula regions. 
This is consistent with the recent work of Rumble et al.~(in prep) that finds that
young stars only provide significant heating of their surroundings when their spectral types
are earlier than B.

Given the similarity in all of these temperature measures, we argue that it is reasonable
to adopt a constant value of 15~K for the bulk of our analysis.

\section{Investigation into CO Contamination of the 850\,$\mu$m Flux in IC\,5146}

A portion of the IC\,5146 Northern Streamer, roughly 0.03 square degrees, was observed in 
$^{12}$CO$(J=3-2)$ using HARP \citep{Buckle10}.  The $^{12}$CO$(J=3-2)$ emission
line lies within the SCUBA-2 850\,$\mu$m bandpass, and can therefore `contaminate'
the thermal dust emission measured \citep{jea03, dea12}.  
Previous observations suggest that
the amount of CO contamination is generally small, typically less than
20\% \citep{jea03,dea12,sea13,Hatchell13,pea15,sea15a,bea15,kea16a,cbea16},
although in zones with weak 850\,$\mu$m emission and a strong CO outflow, CO has been 
observed to contribute up to 90\% of the flux.

We follow the procedure outlined in \citet{dea12} to estimate the level
of CO contamination where suitable observations exist within IC\,5146.
First, the $^{12}$CO$(J=3-2)$ HARP observations were reduced using ORAC-DR \citep{Jenness15}.
Next, we re-ran the SCUBA-2 map-making data reduction process with the integrated CO intensity subtracted 
from the 850\,$\mu$m emission seen in each raw data file, after scaling the CO integrated intensity based on 
the atmospheric transmission at the time of the observation.  Finally, we mosaicked the
CO subtracted maps together and compared the resultant image  with the original 
850\,$\mu$m mosaic.

The fractional CO contamination level is given by
\begin{equation}
f_{CO} = \frac{S_{850,orig}-S_{850,noco}}{S_{850,orig}}
\end{equation}
where $S_{850,orig}$ is the flux in the original 850\,$\mu$m map and 
$S_{850,noco}$ is the flux in the CO-subtracted 850\,$\mu$m map.
Using the clump boundaries identified with FellWalker (\S \ref{sec:anal:clumps}), 
we calculate the
amount of CO contamination for each of the clumps where CO observations
exist over at least part of their extent.  Table \ref{tab:CO} reports the CO contamination
both for the total flux of each clump as well as the peak flux, in addition to
the fraction of the clump area in which CO was observed.
As can be seen from Table \ref{tab:CO}, the CO contamination is never higher than 10\%
and is often closer to 1\%.  For the fainter clumps, the CO contamination level
is smaller than the pixel-to-pixel noise level, as is apparent by several values
of negative contamination reported.

\begin{deluxetable}{ccrr}
\tablecolumns{4}
\tablewidth{0pt}
\tabletypesize{\scriptsize}
\tablecaption{CO Contamination of Clumps\label{tab:CO}}
\tablehead{
\colhead{Clump\tablenotemark{a}}&
\colhead{F$_{area}$\tablenotemark{b}}&
\colhead{f$_{CO,tot}$\tablenotemark{c}}&
\colhead{f$_{CO,pk}$\tablenotemark{c}}
}
\startdata
  25 &  0.301  &  0.020  &  0.000 \\
  26 &  0.743  &  0.060  &  0.074 \\
  27 &  1.000  &  0.031  &  0.028 \\
  28 &  0.026  &  0.001  &  0.000 \\
  30 &  1.000  &  0.043  &  0.061 \\
  31 &  1.000  &  0.031  &  0.030 \\
  32 &  1.000  &  0.017  &  0.025 \\
  33 &  1.000  &  0.017  &  0.004 \\
  34 &  1.000  & -0.020  & -0.004 \\
  35 &  1.000  & -0.003  &  0.018 \\
  36 &  1.000  &  0.019  & -0.000 \\
  38 &  1.000  & -0.011  & -0.022 \\
  40 &  1.000  &  0.011  &  0.022 \\
  42 &  1.000  & -0.008  &  0.005 \\
  43 &  1.000  &  0.036  & -0.000 \\
  45 &  1.000  & -0.016  & -0.005 \\
  46 &  1.000  &  0.005  &  0.004 \\
  47 &  1.000  &  0.069  &  0.045 \\
  48 &  0.399  &  0.026  &  0.006 \\
  50 &  0.782  &  0.094  &  0.049 \\
  52 &  0.897  &  0.031  & -0.002 \\
\enddata
\tablenotetext{a}{Clump designation, as in Table~3.}
\tablenotetext{b}{Fraction of area of clump where CO was mapped.}
\tablenotetext{c}{Fraction of clump's total flux and peak flux attributable to CO contamination.}
\end{deluxetable}

\bibliographystyle{aasjournal}
\bibliography{ic5146}{}

\end{document}